\documentclass[preprint,12pt]{aastex}
\usepackage{lscape}
\usepackage{hyperref}

\def\um{$\mu$m}
\begin{document}
%
\title{Spitzer Observations of Dust Emission from \ion{H}{2} Regions
in the Large Magellanic Cloud}
%
%
\author{Ian W. Stephens\altaffilmark{1,2}, Jessica Marie Evans\altaffilmark{1},
Rui Xue\altaffilmark{1},You-Hua Chu\altaffilmark{1},
Robert A. Gruendl\altaffilmark{1}, and Dominique M. Segura-Cox\altaffilmark{1}}
\altaffiltext{1}{\itshape Department of Astronomy, University of Illinois
at Urbana-Champaign, 1002 West Green Street, Urbana, IL 61801, USA;}
\altaffiltext{2}{\itshape Now at Institute for Astrophysical Research, Boston University, Boston, MA 02215, USA;
ianws@bu.edu}

%
%
%
%
\begin{abstract}

Massive stars can alter physical conditions and properties of their ambient interstellar dust grains via radiative heating and shocks. The \ion{H}{2} regions in the Large Magellanic Cloud (LMC) offer ideal sites to study the stellar energy feedback effects on dust because stars can be resolved, and the galaxy's nearly face-on orientation allows us to unambiguously associate \ion{H}{2} regions with their ionizing massive stars. The \emph{Spitzer Space Telescope} survey of the LMC provides multi-wavelength (3.6--160 $\mu$m) photometric data of all \ion{H}{2} regions. To investigate the evolution of dust properties around massive stars, we have analyzed spatially-resolved IR dust emission from two classical \ion{H}{2} regions (N63 and N180) and two simple superbubbles (N70 and N144) in the LMC. We produce photometric spectral energy distributions (SEDs) of numerous small subregions for each region based on its stellar distributions and nebular morphologies. We use \texttt{DustEM} dust emission model fits to characterize the dust properties. color--color diagrams and model fits are compared with the radiation field (estimated from photometric and spectroscopic surveys). Strong radial variations of SEDs can be seen throughout the regions, reflecting the available radiative heating. Emission from very small grains drastically increases at locations where the radiation field is the highest, while  polycyclic aromatic hydrocarbons (PAHs) appear to be destroyed. PAH emission is the strongest in the presence of molecular clouds, provided that the radiation field is low.

\end{abstract}

\subjectheadings{ISM: dust, extinction -- \ion{H}{2} regions --  infrared: ISM --  Magellanic Clouds}
\maketitle
%
%
%
%
\section{Introduction} Ê\label{sec:intro}

Far infrared (FIR) emission of interstellar dust has been commonly used
to gauge star formation rates \citep[e.g.,][]{Cetal10}. ÊIn a star-forming
region, interstellar dust can be heated, sputtered, and shattered
by energy feedback from massive stars via ultraviolet (UV) radiation,
fast stellar winds, and supernova remnant (SNR) shocks. ÊIt would thus be
prudent to investigate how dust properties are modified in the harsh
environment around massive stars. Ê

The Large Magellanic Cloud (LMC) provides a nearby, ideal laboratory
to study the influence of massive stars on dust properties because
of the following advantages over our Galaxy:
(1) the LMC has a nearly face-on orientation, mitigating the confusion
and extinction along the Galactic plane; (2) the LMC is at a known
distance, $\sim$50 kpc \citep{fea99}, so stars can be resolved
and studied in conjunction with the interstellar gas and dust; and (3)
there exist many complementary multi-wavelength observations of stars,
gas, and dust in the LMC.
For these reasons, the LMC has been targeted by every
space-based IR observatory to study dust properties and calibrate
dust emission as star formation indicators.

The first report of \emph{Spitzer Space Telescope} Êobservations
of an LMC \ion{H}{2} complex was made by \citet{Getal04} for
LHA\,120-N206 \citep[N206 for short; designation from][]{hen56};
its dust emission was
qualitatively compared with that of the Orion Nebula. ÊSubsequently,
the entire LMC was surveyed by \emph{Spitzer} in
the Legacy program Surveying the Agents of a Galaxy's Evolution
\citep[SAGE;][]{Meixner06}. ÊÊUsing the SAGE Êand additional
data of 32 \ion{H}{2} complexes in the LMC and Small Magellanic Cloud (SMC), \citet{Letal10}
found that the 70 $\mu$m fluxes appear to be the best star
formation indicator, although the correlation between the
70 $\mu$m flux and the bolometric IR flux
in the 30 Dor giant \ion{H}{2} region shows a large degree of scatter. ÊThe anomaly in 30 Dor
could be attributed to the intense energy feedback from the
large number of massive stars encompassed by this region.

The SAGE data have also been used to analyze the global FIR emission
in the LMC and spectral energy distributions (SEDs) of three large regions
selected to represent different levels of star-formation activity \citep{Betal08}. A 70 $\mu$m excess in the SEDs is found, and this excess increases
along the sequence ``Milky Way--LMC--SMC'', suggesting an influence of
decreasing metallicity. Ê\citet{Betal08} fit the SEDs with an improved version of
the dust emission model by \citet{Detal90} and suggested that the 70~$\mu$m 
excess can be explained by modifying the size distribution of very small grains (VSGs)
to have an increased abundance of larger VSGs.
Since large VSGs can be produced through erosion of large
grains in the diffuse medium, the presence of VSGs suggests
effects of energy feedback.

To investigate the effects of massive stars on dust emission,
\citet{Slater11} studied the dust properties of 4 classical
and 12 superbubble \ion{H}{2} regions in the LMC and found
little correlation between dust emission properties
and the spectral type of the hottest star, the bolometric luminosity
of the underlying massive stars, or the evolutionary status of
the \ion{H}{2} regions. ÊThis result is surprising and is at odds
with the anomalous properties of 30 Dor and the abundant presence
of large VSGs.

We have examined the \ion{H}{2} region sample of \citet{Slater11}
and find that their division between classical and superbubble
cases may not be consistent with observations. ÊFor example, DEM\,L34 \citep[= N11; DEM
designation from][]{DEM76}
and DEM\,L152 (= N44) both contain multiple OB associations
with a central superbubble surrounded by compact \ion{H}{2} regions,
but are classified as ``classical'' and ``superbubble'' \ion{H}{2}
regions, respectively. ÊFurthermore, they have used the integrated
dust emission from entire \ion{H}{2} regions, while it is expected that
the influence of stellar energy feedback ought to depend
on radial distances from the stars.

To investigate the influence of massive stars on interstellar
dust grains with the least ambiguity, we have chosen to study
two clearly defined classical \ion{H}{2} regions, N63 (=
DEM\,L243) and N180 (= DEM\,L323), and two relatively simple
superbubbles, N70 (= DEM\,L301) and N144 (= DEM\,L199),
whose stellar content is well studied.
We have analyzed \emph{Spitzer} observations of these four
regions, measuring flux densities in radially distributed
subregions, constructing SEDs, and using dust emission model
fits to determine the dust properties. ÊIn this paper, we
describe our methodology, target selection, and data sets used
in Section \ref{sec:hii}, discuss the stellar content and assess the stellar
energy feedback in Section \ref{sec:SCRF}, describe the creation and modeling
of SEDs in Section \ref{sec:SEDs}, examine the results of each \ion{H}{2} region in Section \ref{sec:dust_analysis},
discuss the relationship between stellar energy feedback and dust
properties in Section \ref{sec:disc}, and summarize the findings in Section \ref{sec:sum}.


\section{Overview of Methodology, Targets, and Observations} Ê\label{sec:hii}

Our strategy for studying the effects of stellar energy feedback on dust grains
was to (1) select clear-cut classical and superbubble \ion{H}{2} regions around
OB associations with well-studied stellar content, (2) use the known stellar
content to estimate the radiation field,
(3) use archival \emph{Spitzer} observations to extract SEDs for a series of subregions sampling different radial distances to the ionizing OB association,
(4) use dust emission models to fit the SEDs and assess the dust
properties with respect to the stellar radiation field, and (5) examine the evolutionary status and dust properties of the
\ion{H}{2} regions and search for variations of dust properties that may be
caused by stellar feedback. Ê

Below we describe the four \ion{H}{2} regions selected for this study and
the data sets used in the analysis. ÊThe methods of analysis are described
in detail in later sections.

\subsection{\ion{H}{2}~Regions Selected for Analysis}
We have selected two pairs of \ion{H}{2} regions to study
their spatially resolved dust properties. ÊThe first pair consists of young
 \ion{H}{2} regions, in which massive stars have not cleared
a large central cavity via fast stellar winds and supernova explosions.
The H$\alpha$ surface brightness peaks near the central OB association.
Such \ion{H}{2} regions are often referred to as ``classical'' \ion{H}{2}
regions. For this type we have chosen N63 and N180,
also known as DEM\,L243 and DEM\,L323, respectively.

\begin{figure}
\includegraphics[scale=0.85]{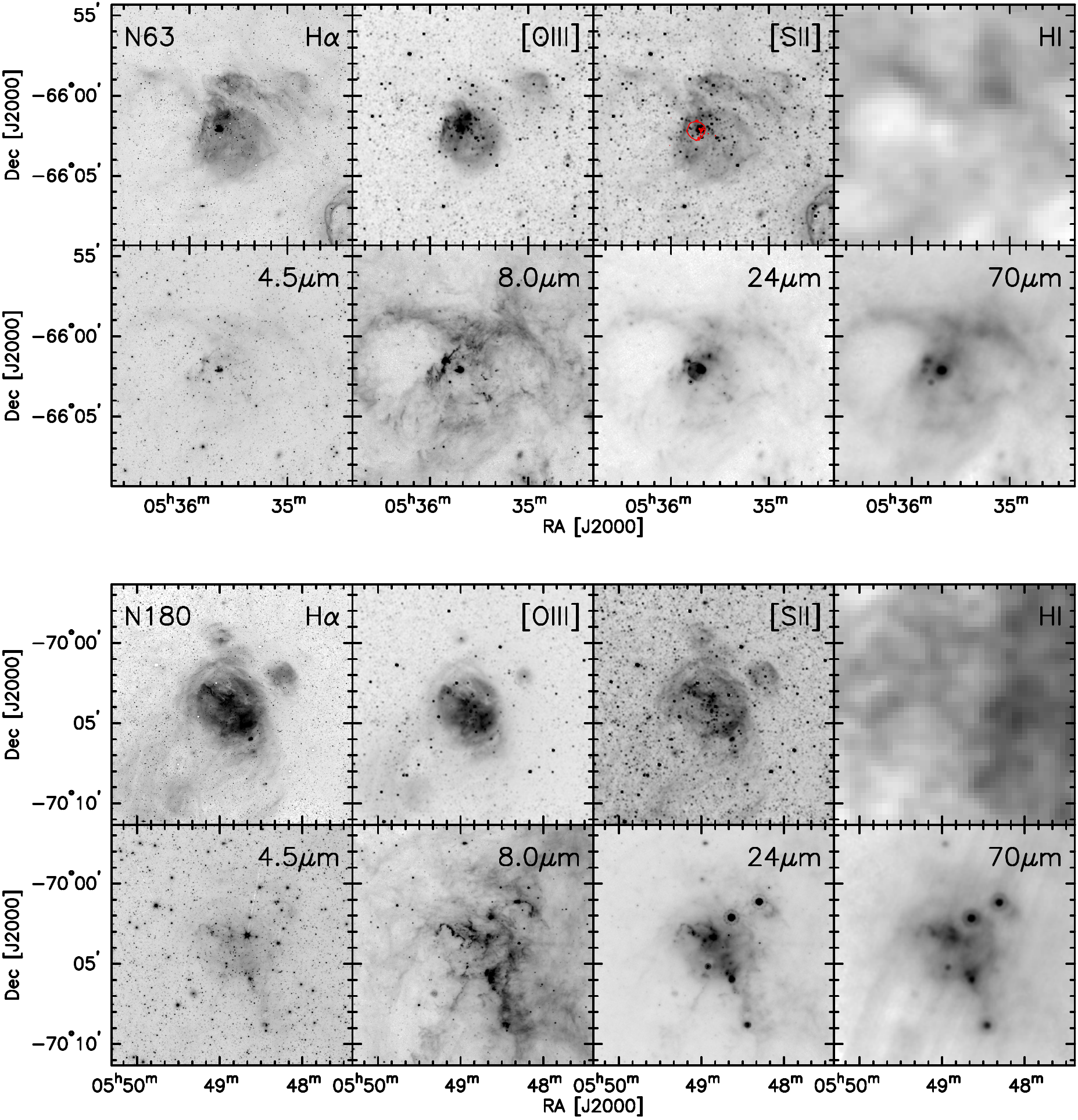}
\caption[Multi-Wavelength Imaging for Classical \ion{H}{2}~Regions N63 and N180]{Multi-wavelength imaging for classical \ion{H}{2} regions N63 and N180. Red contours on [\ion{S}{2}] indicate locations of X-ray emission. \label{four_SB_fig1.ps}}
\end{figure}

{\bf N63} encompasses the OB association LH\,83 \citep{LH70}. ÊThe stellar
content of LH\,83 has been studied both spectroscopically and
photometrically by \citet{Oey96}. ÊAs seen in Figure \ref{four_SB_fig1.ps}, the
\ion{H}{2} region appears amorphous with a few bright patches.
The most massive star in this region has exploded and produced
the SNR N63A \citep{Mathew83,Chu97,Warren03},
marked by the X-ray contour over the [\ion{S}{2}] image in Figure \ref{four_SB_fig1.ps}.
N63 is also classified as a classical \ion{H}{2} region by \citet{Slater11}.

{\bf N180} is photoionized by the OB association LH\,117. ÊThe massive
star content of LH\,117 has been reported by \citet{Massey89}. Ê
Figure \ref{four_SB_fig1.ps} Êshows a bright compact \ion{H}{2} region with some filamentary
structure along its periphery. ÊThe internal kinematics of this
\ion{H}{2} region show local expansion around some massive stars,
but no large-scale expansion or SNR shocks have been
detected \citep{Naze01}. ÊBoth this paper and \citet{Slater11} classify
N180 as a classical \ion{H}{2} region; however, we exclude the faint
southeast extension that is photoionized by the OB association LH\,118 because
the lack of dense gas near the ionizing stars implies a more evolved state.

The second pair consists of superbubbles, in which OB associations
have cleared a central cavity and swept the interstellar gas into a
shell structure. We have selected two superbubbles, N70 and
N144, also known as DEM\,L301 and DEM\,L199, respectively.

{\bf N70} is a superbubble around the OB association LH\,114.
The massive stars in LH\,114 have been studied by \citet{Oey96}.
This superbubble has essentially thermal radio emission (Dopita et
al.\ 1981) but shows detectable diffuse X-ray emission, indicating interactions
with a recent supernova explosion \citep{CM90}. Ê\citet{Slater11} also
classify N70 as a superbubble.

{\bf N144} is a superbubble blown by the OB association LH\,58.
The massive stars of LH\,58 have been studied in detail by
\citet{Garmany94}. ÊWe have classified N144 as a superbubble,
in contrast to the classical \ion{H}{2} region classification
by \citet{Slater11}, because N144 shows a limb-brightened
morphology and an expanding shell structure in high-dispersion,
long-slit echelle spectra in the H$\alpha$ line. ÊNo SNRs have
been reported in N144. ÊMassive young stellar objects (YSOs) have been
found in N144 \citep{GC09}, indicating on-going star formation.

\begin{figure}
\includegraphics[scale=0.85]{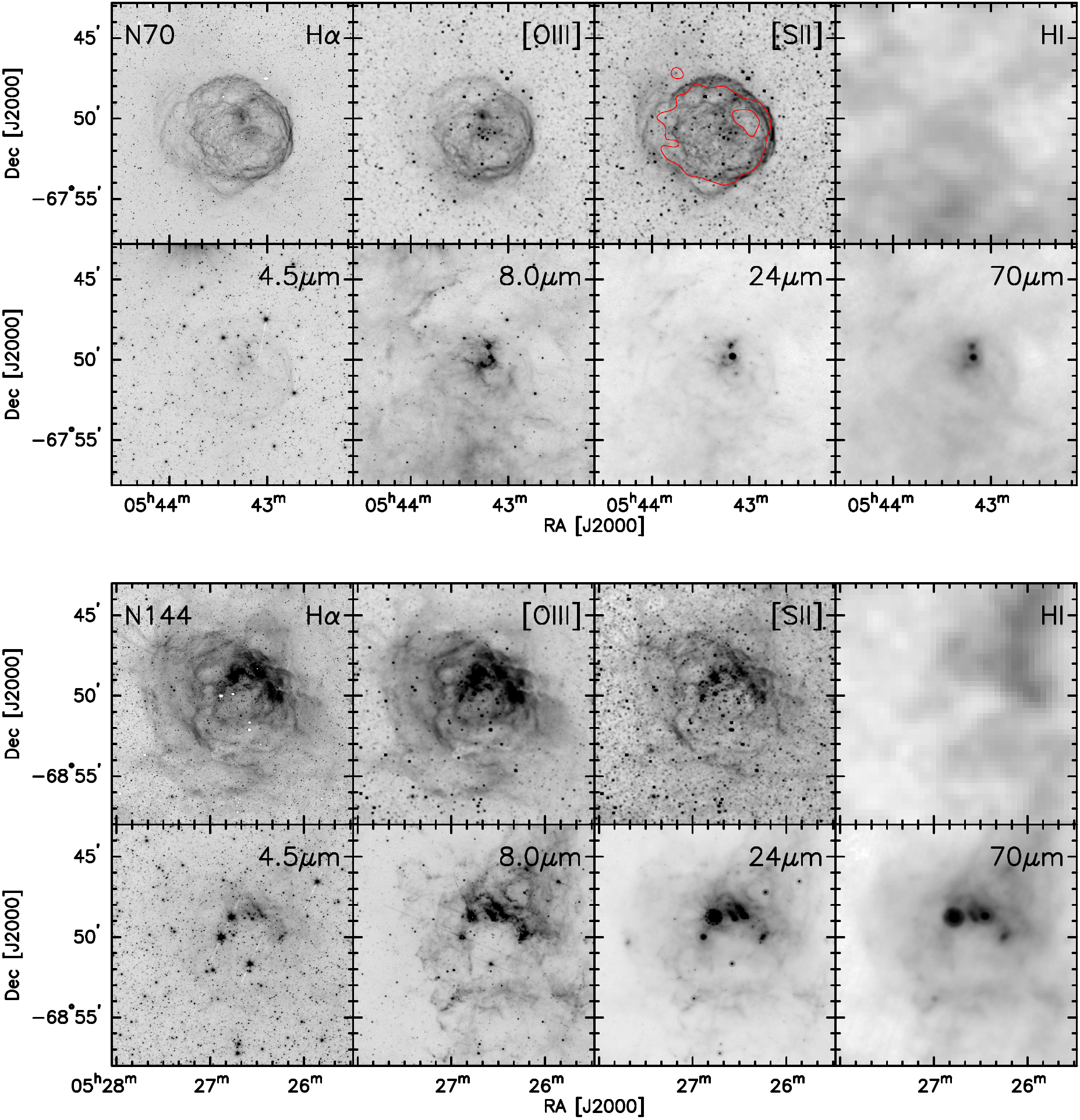}
\caption[Multi-Wavelength Imaging for Supperbubble Regions N70 and N144]{Multi-wavelength imaging for supperbubble regions N70 and N144. Red contours on [\ion{S}{2}] indicate locations of X-ray emission. \label{four_SB_fig2.ps}}
\end{figure}

The emission from stars, gas, and dust in these four \ion{H}{2}
regions can be seen in the multi-wavelength images presented
in Figure \ref{four_SB_fig1.ps} (two classical \ion{H}{2} regions) and Figure \ref{four_SB_fig2.ps}
(two superbubbles). Ê

\subsection{Observations and Data Sets} Ê\label{sec:obs}

Multi-wavelength data sets were used in this work. ÊFirst, we used optical emission-line
images of the \ion{H}{2} regions to examine the spatial distribution and spectral
variations of the ionized gas and to diagnose the ionization fronts and interstellar shocks.
We then used the observed stellar content of these regions to estimate the interstellar
radiation field; we adopted spectroscopically determined stellar types available in the literature, and for the rest we assessed stellar types and masses using optical
photometric data. ÊFinally, we used archival \emph{Spitzer} observations
to extract the SEDs of the dust emission for further modeling and analysis. Ê
These data sets are described below.

\subsubsection{Nebular Emission-line Images}\label{sec:Ha_data}

H$\alpha$, [\ion{O}{3}] $\lambda$5007, and [\ion{S}{2}] $\lambda\lambda$6716, 6731
images of these \ion{H}{2} regions are available from the Magellanic Cloud
Emission-Line Survey \citep[MCELS;][]{Smith99}. Ê
These images are neither continuum-subtracted nor flux-calibrated;
thus, only variations in the surface brightness profiles or line ratios can be
determined quantitatively. ÊA rough H$\alpha$ flux calibration can be made by
comparisons with the integrated H$\alpha$ fluxes reported by \citet{KH86}.
The angular resolution of the MCELS images is 2-3$\arcsec$.

We have obtained higher-resolution H$\alpha$ images of our target \ion{H}{2}
regions using the MOSAIC2 camera on the Blanco 4m telescope at the
Cerro Tololo Inter-American Observatory. ÊThe images were bias and flat-field
corrected. ÊAstrometric solutions were obtained through comparisons with the Two
Micron All Sky Survey Point Source Catalog, typically resulting in better than
$\sim$0\farcs2 accuracy. ÊThe H$\alpha$ images presented in Figures \ref{four_SB_fig1.ps} and \ref{four_SB_fig2.ps}
have $\sim$1\arcsec\ resolution and are combinations of at least
3$\times$300 s exposures. ÊThese MOSAIC H$\alpha$ Êimages are also not
continuum-subtracted or flux-calibrated, but they are useful in showing
positions of ionizing stars, dust lanes, and small-scale structure of the ionized
gas.

\subsubsection{Spectroscopic Classifications of Massive Stars}\label{sec:replstars}

The most massive stars are also the most luminous stars and thus dominate
the radiation field of an OB association. ÊSince hot massive stars emit mostly in
the far-ultraviolet, the effective temperature and luminosity of a massive star
cannot be determined reliably from optical photometry; spectroscopic classifications
are needed \citep{Massey93}. ÊAll four \ion{H}{2} regions selected for this
work have their stellar content well studied both photometrically and spectroscopically.

For the most massive stars, we relied on the spectroscopic classifications published in the literature and have included the following in our sample: for N63, \citet{Oey96} classified 14 OB stars and 2 late-type supergiants, and for N70, classified 18 OB stars and 5 late-type supergiants.  For N144, \citet{Garmany94} listed new and published spectral types for 39 OB stars and 3 late-type supergiants, and for N180 \citet{Massey89} provided new and previously published spectral types for the 19 bluest stars and 2 supergiants.


Ê
The spectral types and luminosity classes of OB stars from these studies were used to determine their masses and contributions to the radiation field. ÊThe late-type supergiants are as luminous as main-sequence B0 stars and were thus included in the calculation of radiation fields.

\subsubsection{Magellanic Clouds Photometric Survey}\label{sec:MCPS}
For main-sequence stars later than late-O, the spectral types can be assessed from optical photometry; therefore, we used the Magellanic Clouds Photometric Survey (MCPS) catalog for stars in the LMC \citep{zar04}. ÊThis catalog was also used to identify stars in areas around the \ion{H}{2} regions which are not covered by previous spectroscopic studies but whose massive stars may make significant contributions to the radiation field. ÊMCPS provides astrometry and $UBVI$ photometry of stars and is reasonably complete for sources with $V \lesssim$ 20~mag. ÊEach MCPS entry contains a set of flags that indicate the quality of the individual detections, including how well the source compares to the point spread function (PSF) and whether the measurements were replaced by photometry from the literature.

We have excluded catalog entries that do not have a complete set of photometric measurements in the $UBV$ bands, whose magnitude errors in any band exceed 0.2 mag, and whose stars are excluded from PSF fitting or are not well fit by the PSF. ÊFurthermore, we excluded detections whose $(B-V)$ $\ge$ 0.2~mag since the $B$ versus $B-V$ color--magnitude diagram (CMD) becomes noisy/random for these values. In these \ion{H}{2} regions, the extinction $E(B-V)$ of massive stars based on spectroscopic studies \citep{Massey89,Garmany94,Oey96} are rarely above 0.25 (with the max at $\sim$0.4). Given the intrinsic $B-V$ values for stars above 7~$M_{\sun}$ \citep[e.g.,][]{allenistheman}, this sample will rarely exclude stars that add significant radiation (see Section \ref{sec:mass_est}) within our \ion{H}{2} regions.




The MCPS photometry in the central region of N180 appeared uncharacteristically erroneous upon initial inspection. ÊThe MCPS observations were made using a drift-scan camera, which was unable to resolve densely distributed stars against a bright nebular background \citep{zar04}. ÊFor this reason, in the high surface brightness central region of N180, we opted for the photometric results from \citet{Massey89}, prioritizing with spectral types from the same study when possible. ÊIn this region, \citet{Massey89} provides 192 photometric observations with $(B-V)$ $<$ 0.2, and 21 spectroscopic observations.



\subsubsection{Spitzer Space Telescope Observations}\label{sec:SAGE_data} ÊÊÊÊÊÊÊÊÊÊÊ Ê
The \emph{Spitzer} SAGE observations of the LMC used the Infrared Array Camera (IRAC) and the Multiband Imaging Photometer (MIPS) instruments. The IRAC images have four bands centered at 3.6, 4.5, 5.8, and 8 $\mu$m, and the MIPS images have three bands centered at 24, 70, and 160 $\mu$m. The detailed survey strategy and data processing of the SAGE observations can be found in \citet{Meixner06}. ÊFor all four IRAC bands, we sampled available 1\fdg1$\times$1\fdg1 tiles of our \ion{H}{2} regions from the most recent SAGE data release, DR3 (2009). ÊFor MIPS, we sampled data from DR3 or DR2 (2008) depending on the availability of specific data. ÊFor example, a complete mosaic for MIPS 160 $\mu$m was available in DR2, but not DR3. ÊIn addition, the available 24 $\mu$m mosaic of the LMC from DR3 was prohibitively large due to its higher resolution, but 1\fdg1$\times$1\fdg1 tiles are not released in DR3; thus the small 24 $\mu$m files from DR2 were used. ÊImage size was not an issue for the 70 $\mu$m mosaic, so the DR3 mosaic was used.

\section{Stellar Content and Radiation Field} Ê\label{sec:SCRF}

To investigate the effects massive stars have on the dust in these \ion{H}{2} regions, we must quantify the stellar radiation field. ÊWe accomplish this by (1) estimating luminosities from their spectral types, when available, or through comparisons of their locations with stellar evolutionary tracks in CMDs, then (2) determining their bolometric incidence flux on designated subregions.

\subsection{Photometric Estimation of Stellar Masses and Luminosities} \label{sec:mass_est}

To estimate the masses and luminosities of stars without spectroscopic classifications, we used the photometric measurements from MCPS and \citet{Massey89} and compared them against stellar evolutionary tracks from \citet{Geneva01} for initial masses between 5 and 120 $M_{\odot}$.
Along these evolutionary tracks, the effective mass, luminosity, temperature, and predicted $UBVRI$ photometry are tabulated for time steps of $\Delta$ log $t = 0.05$ dex starting at 10$^3$~yr. We adopted the evolutionary tracks for a metallicity $Z$ = 0.4~$Z_{\odot}$, which is close to the LMC metallicity.

The extinction to individual stars is unknown, but the stars of interest are blue; therefore, we opted to use the Wesenheit extinction-free magnitude $W Ê\equiv V -$ DM $- (B-V) A_V / E(B-V) $ \citep{Madore82} and the Johnson $Q$ reddening-free color $Q \equiv (U-B) - (B-V) E(U-B) / E(B-V)$ \citep{Johnson53}, where DM = 18.5 is the distance modulus of the LMC, Ê$E(U-B) / E(B-V) = 0.72$~mag for blue stars, and $A_V / E(B-V) = 3.1$~mag from the canonical extinction law.

We assigned masses and luminosities to our candidates by plotting their extinction-free parameters ($W$ and $Q$) in a CMD alongside the evolutionary tracks. ÊAs with the MCPS data, the evolutionary tracks were expressed in extinction-free parameters for direct comparison. ÊThe evolutionary tracks represent stars of masses 5, 7, 10, 12, 15, 20, 25, 40, 60, 85, and 120 $M_{\odot}$. The closest evolutionary track to a star gives an estimate for its mass, and its position along the track determines its luminosity. We remove stars lower than 7~$M_{\odot}$ by matching them to the 5~$M_{\odot}$ track. Our final analysis includes stars with estimated masses $\gtrsim$ 7 $M_{\odot}$. The exclusion of lower mass stars is justified because for an unevolved population with the \citet{Salpeter55} initial mass function (IMF), only $\sim$4\% of the total radiation is contributed by stars $<$ 7 $M_{\odot}$. Indeed, spectroscopic surveys \citep{Massey89,Garmany94,Oey96} have shown that our regions have IMF slopes similar to or steeper than the slope of the \citet{Salpeter55} IMF. Moreover, these surveys show that our regions are relatively unevolved since they still have massive O-stars (O3-4 in N70, N180, and N144 and an O7 in N63). 

\begin{figure}
\center
\includegraphics[angle=-90,scale=.5]{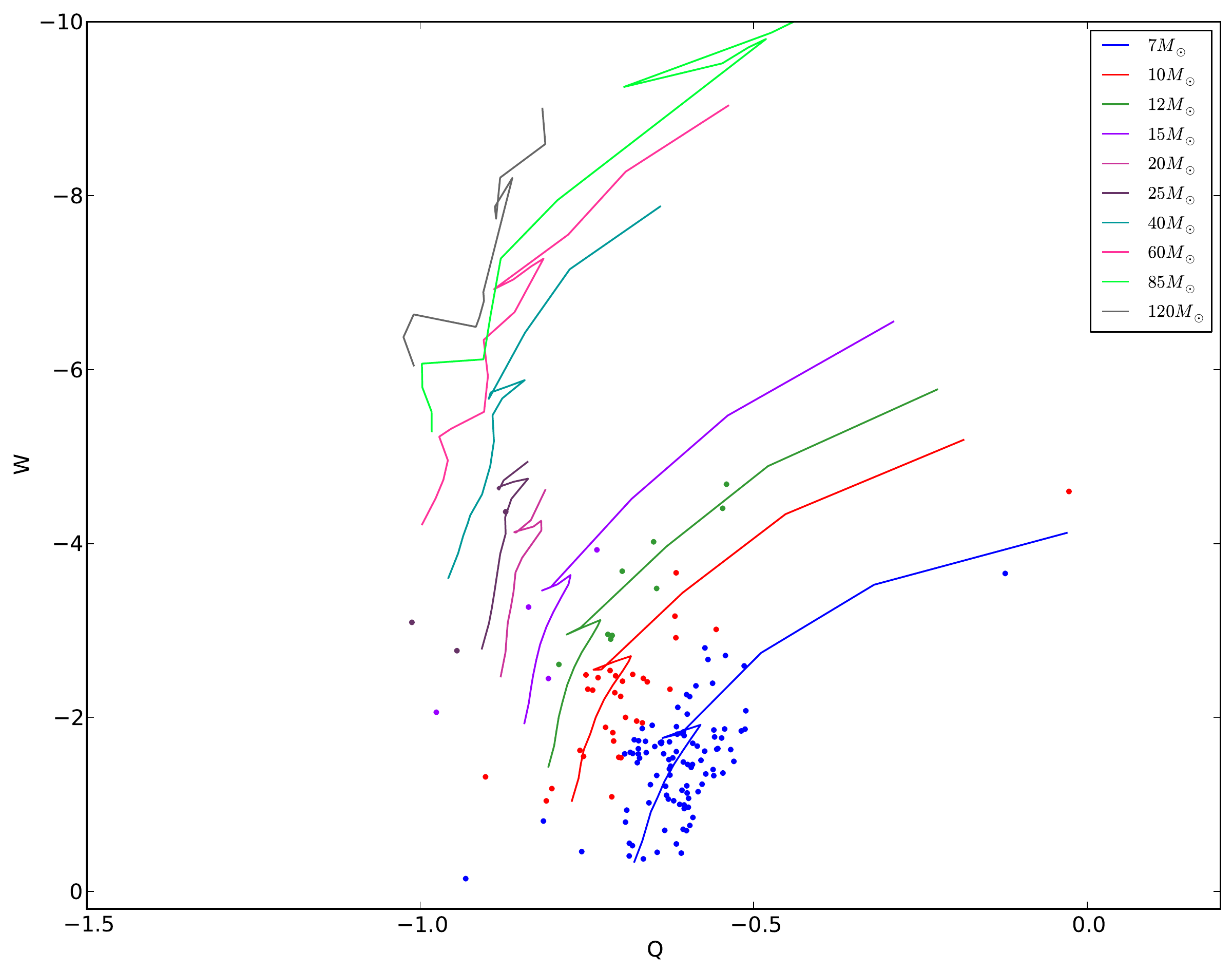}
\caption[Color--Magnitude Diagram of Classical \ion{H}{2}~Region N63 with Evolutionary Tracks]{Color--magnitude diagram of classical \ion{H}{2} region N63 with evolutionary tracks from \citet{Geneva01}. The color of each star matches the color of the closest evolutionary track, indicating its mass. The position along the track provides the luminosity used in the construction of the radiation field. Ê\label{N63_CMD}}
\end{figure}

A sample CMD can be found in Figure \ref{N63_CMD}, which shows evolutionary tracks and stars in N63; the rest of the CMDs are available in the online version of this paper. ÊWith approximated luminosities for the massive star candidates, the radiation field could be quantified. For both N63 and N70, the highest mass stars as ascertained by the evolutionary tracks (and have no spectroscopic classification) were 25~$M_\sun$. N144 and N180, on the other hand, included several stars at 40 and 60~$M_\sun$ that were not observed in spectroscopic surveys. The masses of these stars are uncertain, but this should not affect the calculation of the radiation field incident on our subregions (described in the next section) by more than a factor of two.


\subsection{Radiation Field}\label{sec:rad}


Based on the fitted evolutionary tracks, 146, 208, 682, and 345 massive stars ($M_{\odot}$ $\gtrsim$ 7) determined from photometry exist across the \ion{H}{2} regions N63, N70, N144, and N180, respectively.  The luminosities of these stars were calculated from their positions along their mass evolutionary tracks.  In addition to the photometric sample, 16, 23, 42, and 21 spectroscopic sources exist in N63, N70, N144, and N180, respectively, whose masses and luminosities were estimated directly from their spectral types \citep{johnson66,DJ87,allenistheman,smi02}. Both photometric and spectroscopic luminosities were used in the radiation field calculation.

The line-of-sight locations of the stars are unknown, so it is impossible to compute the three-dimensional radiation field.  As an approximation,  we produced a two-dimensional radiation field image by adding the fluxes of all the stars about each \ion{H}{2} region using an inverse square law with effective distance $d_{\mathrm{eff}} = \sqrt{2}d_{\mathrm{proj}}$, where $d_{\mathrm{proj}}$ is the projected distance to a star. ÊThese images are shown in the rightmost panels of Figures \ref{N63_all}(a)--\ref{N144_all}(a). The radiation flux for various parts of each \ion{H}{2} can be calculated to approximate the total incident power on each subregion for comparison with exhibited dust properties. The calculated interstellar radiation field (ISRF) presented for each subregion is the median ISRF, which allows direct comparison to the methodology discussed in Section \ref{sec:SEDs}.


\section{Constructing and Modeling of Spectral Energy Distributions of Dust Emission} Ê\label{sec:SEDs}
\subsection{Construction of SEDs and color--color Diagrams}

In order to investigate how spectral properties of dust emission vary radially throughout
each \ion{H}{2} region, we (1) defined radial subregions based on ionized features in H$\alpha$ images, (2) filtered out contamination due to individual stars in the IRAC
bands, (3) calculated background-subtracted fluxes in each subregion, (4) established
SEDs for each subregion for further modeling, and (5) constructed color--color diagrams (CCDs) for comparisons among the subregions.

For each \ion{H}{2} region (N63, N180, N70, and N144), we used MOSAIC H$\alpha$ images to examine the distribution of ionized gas in the region. We then placed circular apertures (subregions) with a 40\arcsec-diameter (10 pc at the distance of the LMC) in a pattern bisecting the stellar content to sample the dust emission radially for each \ion{H}{2} region. In N63, N70, and N144 we used a cross pattern (evenly spaced in each of the four cardinal directions), and for N180 we used a single cut from the northeast to the southwest in order to avoid the diffuse \ion{H}{2} region around the OB association LH118 to the southeast. ÊFor each \ion{H}{2} region, the subregions extend beyond the bright H$\alpha$ emission. ÊTo estimate the local foreground and background emission, we placed an additional subregion well outside the ionized region at a location with low flux in all the \emph{Spitzer} bands and devoid of massive stars as
identified in Section \ref{sec:mass_est}. ÊThe locations of the subregions are marked in the top middle panel of Figures \ref{N63_all}-\ref{N144_all}.

To reduce the contributions of bright stars to the IR emission, we used a box median filter with size 4\farcs2$\times$4\farcs2 (7$\times$7 pixels) to remove point sources from images in the IRAC bands. ÊIn the median-filtered images, several of the brightest remaining objects can be identified as extended background galaxies. As a test, SEDs were computed for subregions using both median-filtered and unaltered images for N63. The uncertainty estimates (the method of calculating these are discussed below) indeed show $\sim$25\% improvement
when the contamination by individual stars has been median-filtered out.
The MIPS images were not median-filtered, as stellar emission is negligible compared to dust emission at these long wavelengths.

The flux density of each subregion was calculated by measuring the median flux value of the pixels within each subregion (using the median-filtered IRAC images and unaltered MIPS images) and multiplying by the total pixel count. ÊA background subregion's flux density was then subtracted from each subregion's flux density to determine the local emission. The median was chosen rather than the mean because large ranges of fluxes with non-Gaussian distributions were often present. ÊTo assess the approximate uncertainty of each measurement, we used the first and third quartile of the flux distribution within the subregion as estimates for the lower and upper uncertainty estimates, respectively. ÊSince the resolution and data-sampling of \emph{Spitzer} is wavelength-dependent, the number of pixels which formed each flux measurement varies with band. A subregion in each of the four IRAC bands contained 3409 pixels, while the MIPS 24, 70, and 160~$\mu$m bands had 197, 49, and 5 pixels, respectively. Therefore, the quartile error-bars show the spread of subregional fluxes for all bands except MIPS 160~$\mu$m (however the 160~$\mu$m error bars do not grossly affect SED fitting). The flux densities from the seven $Spitzer$ mid-IR bands were then combined to form an SED for each subregion. The photometric flux density measurements of each subregion for all four \ion{H}{2} regions are shown in Table \ref{tab_all_fl}.


Our SEDs, as derived from median-filtered images, were also compared to SEDs made from SAGE's smoothed and point-source-subtracted residual images \citep{Meixner06}. Subregion fluxes for the median-filtered images typically differed from the residual images by less than 10\%, though for N144, which has the most stars, fluxes differed by about 15\%. Therefore, the general shape of the SEDs for both the median-filtered images and SAGE residual images were very similar. The point-source subtractions in the SAGE residual images sometimes caused artifacts within the subregions and were often either under- or over-subtracted. For these reasons, we preferred our median-filtered images, though similar results would be obtained if we used the residual images.

Finally, we used these photometric measurements to construct CCDs for each \ion{H}{2} region to search for SED slope changes and color groupings among the subregions. The color magnitudes were calculated based on the Vega zero-magnitude flux densities: 280.9 $\pm$ 4.1, 179.7 $\pm$ 2.6, 115.0 $\pm$ 1.7, and 64.13 $\pm$ 0.94 Jy for the 3.6, 4.5, 5.8, and 8.0 $\mu$m bands respectively \citep{Reach05} and 7.17 $\pm$ 0.11 Jy for 24 $\mu$m \citep{Rieke08}. CCDs including 70 and 160 $\mu$m measurements were not included due to lack of distinct spectral features seen in SEDs, low pixel count sampled, and lack of zero-magnitude flux at these wavelengths in the literature. CCDs for each regions can be seen in panels (c) and (d) of Figures \ref{N63_all}-\ref{N144_all}.

\subsection{SED Modeling}

In order to study the dust properties in each subregion, we simulated the SEDs using the dust emission modeling tool \texttt{DustEM}\footnote{\url{http://www.ias.u-psud.fr/DustEM/}} \citep{com2011}. The \texttt{DustEM} code allows an arbitrary combination of various grain types as input and produces a model SED based on the incident radiation field strength and grain physics in the optically thin limit. We adopted the \citet{Detal90} dust model implemented in \texttt{DustEM} which includes stochastic heating of grains and consists of three dust grain populations:  polycyclic aromatic hydrocarbons (PAHs), VSGs composed of carbonaceous material, and big grains (BGs) made of astronomical silicates. The free parameters in the SED fitting include the mass column density of each dust population and the ISRF parameter $U$, as scaled from the solar neighborhood ISRF described by \citet{Mathis83}. With $U$ as a free parameter rather than temperature, the model inherently assumes that radiation is the dominant component for heating the dust grains instead of other processes (e.g., shocks), which is likely the case for our \ion{H}{2} regions.

The model SEDs were integrated over the filter transmission\footnote{\url{http://www.astro.caltech.edu/~capak/cosmos/filters/}} of each band to get photometric fluxes. ÊThe model fluxes were compared with the observed fluxes, and the smallest $\chi^2$ value was used to determine the best-fit model.
As the 3.6 and 4.5~$\mu$m flux densities are potentially contaminated by starlight, hydrogen Br$\alpha$ line, and/or molecular hydrogen lines \citep{flag2006,church2004,DeBu10}, they were excluded in the SED fitting (more on this below). We note that the value found for $U$ is primarily constrained by the band flux ratio of 70 and 160~$\mu$m, and the PAHs and VSG mass column densities are most sensitive to IRAC bands and 24~$\mu$m band, respectively. 

Table \ref{tab_dustem} summarizes the modeling results. The best-fit dust composition is given as $Y_{\rm PAH}/Y_{\rm DUST}$ and $Y_{\rm VSG}/Y_{\rm BG}$, in which $Y$ is the dust mass abundance for a given grain type. The visual extinction, $A_V$, was derived from the modeled extinction curve and is an indication of the dust column density. The median and range of the BG equilibrium temperatures, $T_{\rm{eq}}$, are also listed in the table and are highly dependent on the value of the modeled $U$. $T_{\rm{eq}}$ values in each region typically peak toward the center and decreases radially. We do not list the fits for a subregion in the following situations: (1) if one of the 70~$\mu$m and 160~$\mu$m bands used for a fit has a negative background subtracted flux ($U$ strongly depends on these bands, and, without $U$, the other free parameters usually are not well-constrained), (2) at least three of the four fitted parameters are less than or equal to their corresponding error bars, and (3) the case of N70 subregion 6, which had an unphysical ISRF fit of $U$ = 285.0$\pm$69.0 due to a large 160~$\mu$m error bar.

\begin{figure}
\center
\includegraphics[scale=.7]{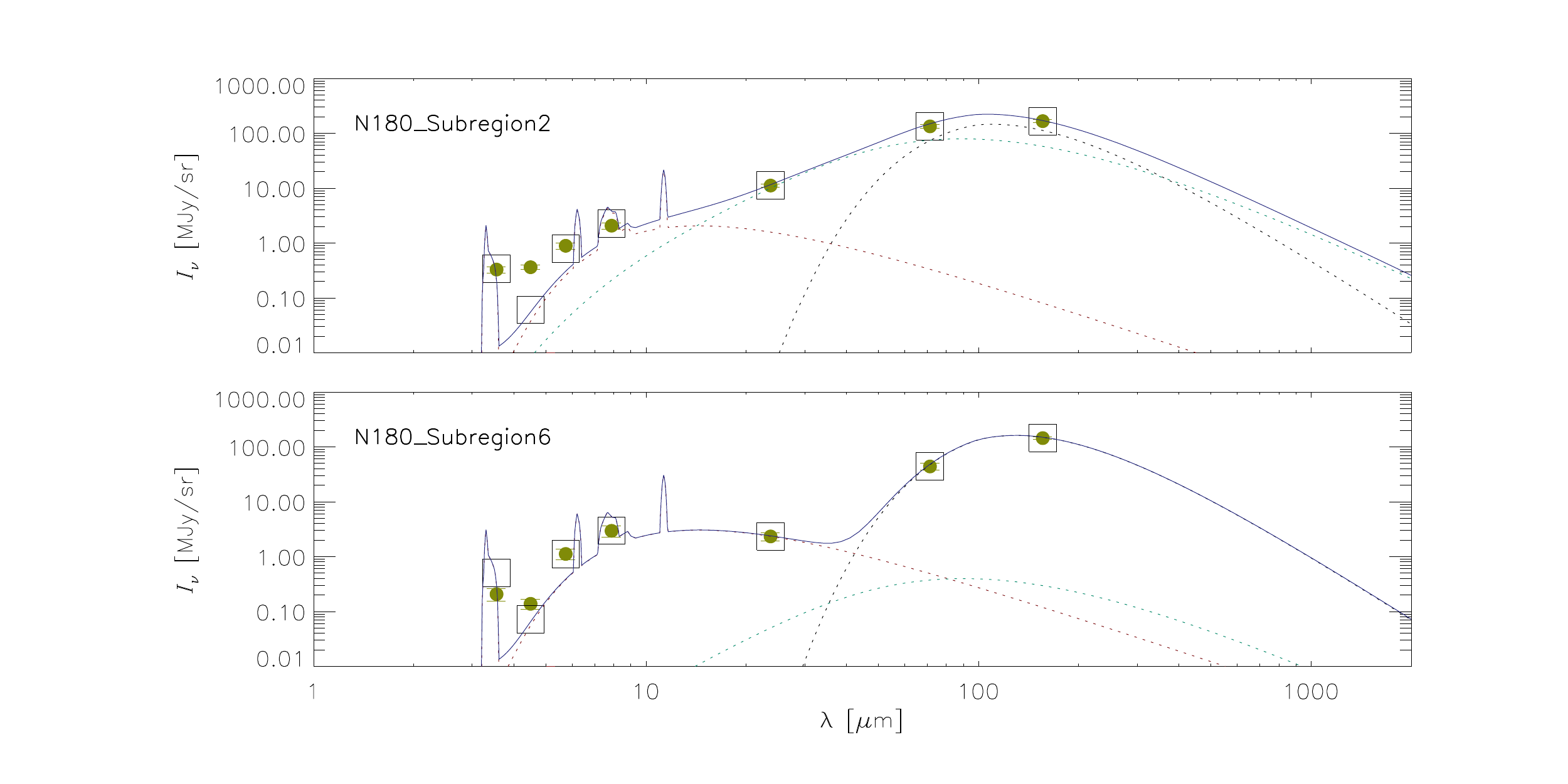}
\caption[Example \texttt{DustEM} Fits]{Examples of \texttt{DustEM} SED fits for subregions in N180. We present a VSG dominated subregion (top panel) and a PAH dominated subregion (bottom panel).  Contributions from each of the three grains are shown with dashed lines in different colors, with maroon showing PAHs, teal showing VSGs, and black showing BGs. The sum of the emission of the three dust components is shown by the solid blue line. Yellow points and error bars show points of the IRAC/MIPS photometric SED. Squares show the fitted locations at the wavelength of each IRAC/MIPS band.\label{fig_n180sed}}
\end{figure}

An example of the dust emission model is shown in Figure \ref{fig_n180sed} for N180 subregion 2 (high $Y_{\rm VSG}/Y_{\rm BG}$) and subregion 6 (high $Y_{\rm PAH}/Y_{\rm DUST}$). The model emission contributed by three grain populations are presented in dashed or dotted lines, with the total emission in the solid blue line. While the model SED reproduced the photometric flux densities included in the SED fitting, it did not predict 3.6 and 4.5~$\mu$m well. In fact, for most locations of our four regions, the 4.5~$\mu$m dust emission produced by the \citet{Detal90} model is much lower than observed. To explain the 3.6 and 4.5~$\mu$m Êdiscrepancy, we tested more complicated PAHs models from \citet{draine2007} and \citet{com2011} (also implemented in \texttt{DustEM}), which have both neutral and ionized PAHs. Although the usual small discrepancy (less than a factor of 3) at 3.6~$\mu$m can be explained by adjusting the PAH ionization fraction, the 4.5~$\mu$m dust model flux is still underpredicted. Thus, for our dust modeling, we excluded the 3.6 and 4.5~$\mu$m bands from our SED fits. Due to the lack of 4-5~$\mu$m spectroscopic observations to resolve the line and continuum emission, any further interpretation is beyond the scope of this work. Therefore, we have five fixed parameters (two IRAC and three MIPS bands) and four fitted parameters ($Y_{\rm PAH}$, $Y_{\rm VSG}$, $Y_{\rm BG}$, and $U$). 

We note that the \citet{Detal90} modeling of VSGs differs from many recent dust models \citep[e.g.,][]{draine2007,com10}. VSGs in newer models are typically modeled by multiple distributions of different grain types with sizes ranging from $\sim$3-400~\AA. However, due to the limited number of bands, our data are best modeled by one VSG component, which, in the \citet{Detal90} model, consists of ``medium-sized" VSGs that have radii between 12 and 150~\AA. This grain is highly dependent on whether or not an SED has a 24~$\mu$m excess. With only three grain types, we also note that BG and VSG are typically anti-correlated; thus, the change in the fraction of $Y_{\rm VSG}/Y_{\rm BG}$ may be somewhat over-exaggerated. 


\section{Analysis}\label{sec:dust_analysis}


For each \ion{H}{2} region we have utilized photometric and spectroscopic data of massive stars throughout the region to estimate the radiation field, constructed and fitted SEDs for the radially placed subregions, and created CCDs for the subregions. The combination of the results are presented for each region in Figures \ref{N63_all}--\ref{N144_all}.

For each figure, panel (a) shows the region as observed in the H$\alpha$ line and the IRAC 8.0~$\mu$m band as well as the ISRF calculated from stars (Section 3.2). The sampled subregions are shown on top of the 8.0~$\mu$m image. Contours of \ion{H}{1} column density are shown in red and CO column density in blue over the H$\alpha$ image.
The radiation field can be compared to the morphology of the \ion{H}{2} region to assess radial variations of dust properties due to massive stars.

Panel (b) shows the results of the best-fit \texttt{DustEM} model for the SED of each subregion as a function of offset from the central subregion.  These fitted parameters include the visual extinction ($A_V$), fitted ISRF (with magenta and cyan colors to show the ISRF as derived from stars for comparison), the ratio of PAH mass abundance to total dust mass abundance ($Y_{\rm PAH}$/$Y_{\rm DUST}$), and the ratio of VSG mass abundance to BG mass abundance ($Y_{\rm VSG}$/$Y_{\rm BG}$). Subregions with photometric measurements approximately equal to the background are excluded from the panel.

For panels (c) and (d) we show [3.6]--[4.5] versus [5.8]--[8.0] and [4.5]--[5.8] versus [8.0]--[24] CCDs, respectively. Note that IRAC bands 1 (3.6 $\mu$m), 3 (5.8 $\mu$m), and 4 Ê(8.0 $\mu$m) contain PAH emission, while IRAC band 2 (4.5 $\mu$m) and the MIPS 24 $\mu$m band do not. Therefore, in our CCDs, a subregion with a higher PAH mass fraction (i.e., higher fractional PAH emission) should have a smaller [3.6]--[4.5], a larger [4.5]--[5.8], and a lower
[8.0]--[24]. In other words, we expect subregions with an increased PAH mass fraction to be toward the bottom of the [3.6]--[4.5] versus [5.8]--[8.0] CCD and toward the top left for the [4.5]--[5.8] versus [8.0]--[24] CCD. Subregions that are not present in these CCDs are those whose surface brightnesses are low and approach that of the background subregion for the \ion{H}{2} region.

\subsection{Classical \ion{H}{2}~Regions--N63, N180}

\subsubsection{N63}
\begin{figure} Ê
\begin{center}
\includegraphics[scale=0.75]{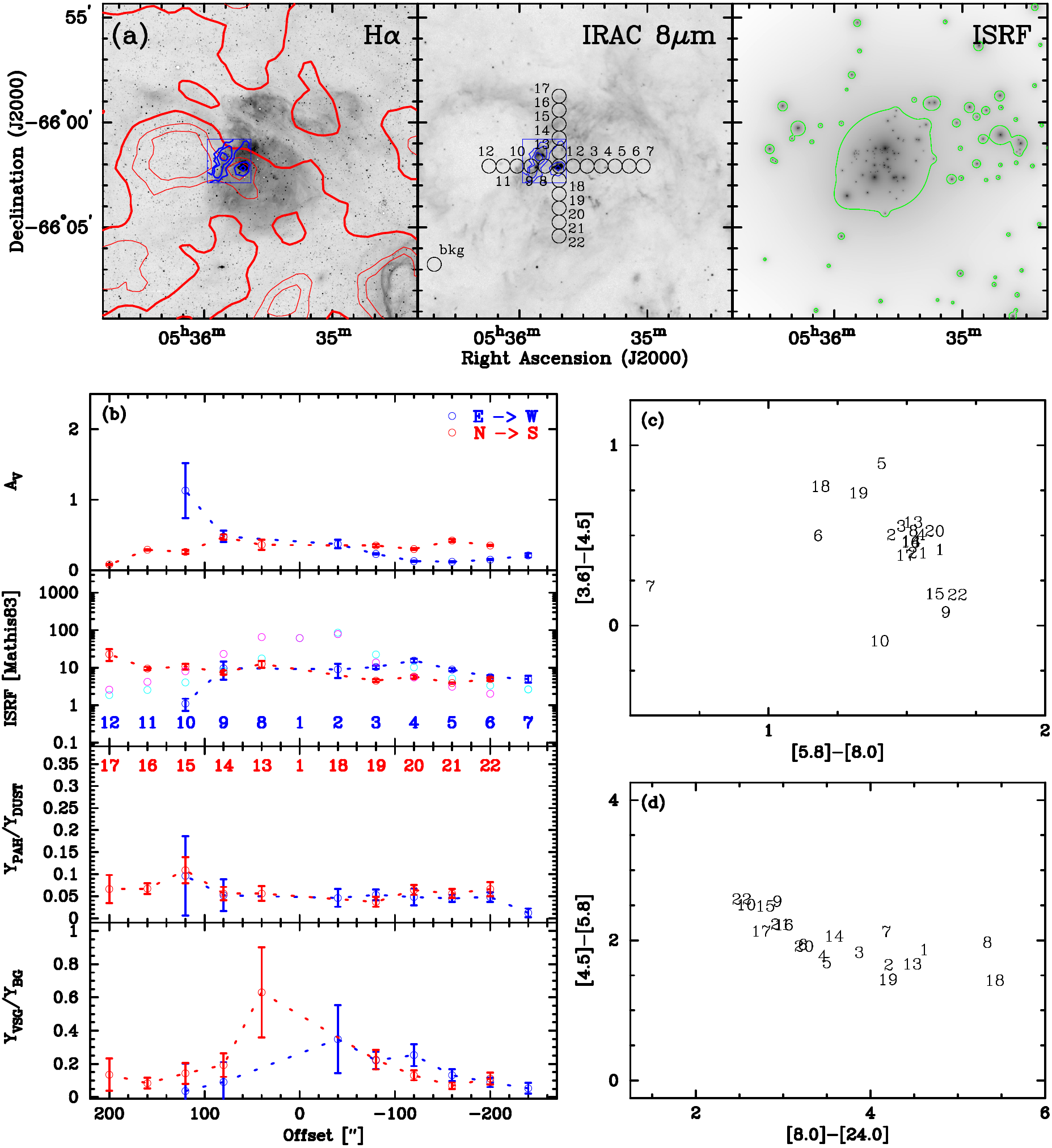}
\caption[Observations and Results of Classical \ion{H}{2}~Region N63]{Observations and results of classical \ion{H}{2} region N63. (a) H$\alpha$, IRAC 8 $\mu$m, and the radiation field due to massive ($\gtrsim7M_{\sun}$) stars. North is up. Red contours show ATCA+Parkes \ion{H}{1} integrated intensity contours. Blue contours are ESO-SEST CO(2--1) integrated intensity at 22$\arcsec$ resolution. Green contours on the rightmost panel indicate a fiducial value for the radiation field due to massive stars. Numbered circles show 40$\arcsec$ subregions sampled with the background subregion labeled as ÒbkgÓ. (b) \texttt{DustEM} model fits. The x--axis shows the offset from the central subregion. The variations of the physical quantities from top panel to bottom panel are: visual extinction, InterStellar Radiation Field, PAH mass fraction to dust mass fraction, and VSG mass fraction to BG mass fraction.  The magenta and cyan open circles on the ISRF panel show the ISRF as derived from massive stars (Section \ref{sec:rad}). Error bars are 1$\sigma$. (c) [3.6]--[4.5] versus [5.8]--[8.0] and (d) [4.5]--[5.8] versus [8.0]--[24.0] color--color diagrams. \label{N63_all}
}
\end{center}
\end{figure}

The classical \ion{H}{2} region N63 has a central concentration of massive stars surrounded by a scattered distribution. The ISRF is centrally peaked at subregions 1, 2, 13, 18, and 19 (Figure \ref{N63_all}). The SNR N63A (marked by X-ray contours in Figure \ref{four_SB_fig1.ps}) dominates emission at 24 $\mu$m (see Figure \ref{four_SB_fig2.ps}) in subregions 1, 8, and 18 due to the heated dust emission and the [\ion{O}{4}] 25.89 $\mu$m line \citep[][D. Segura-Cox et al. in preparation]{Caulet12}; thus, locations of these subregions on the [4.5]--[5.8] versus [8.0]--[24.0] CCD and the \texttt{DustEM} fits should be interpreted with caution. Large-scale molecular clouds as surveyed by NANTEN \citep{Fukui08} were undetected. However, supplemental ESO-SEST CO observations found molecular emission and are shown in Figure \ref{N63_all}(a) (with the region observed outlined in blue, D. Segura-Cox et al. in preparation). ÊWe note that the 8.0 $\mu$m emission and the CO emission are coincident with each other.

For the CCD in Figure \ref{N63_all}(c), we have three subregion groups (ignoring the faint emission from subregion 7): (1) 9, 15, and 22, (2) 5, 6, 18, and 19, and (3) all others. Group 1 has the most fractional PAH emission; subregion 9 is contained within a molecular cloud, subregion 15 straddles the boundary between ionized gas and a dust lane, and subregion 22 is located just outside the \ion{H}{2} region. ÊThese environments are where photodissociation regions (PDRs) are expected; thus, their high fractional PAH emission is not surprising. Group 2 indicates the least fractional PAH emission. Subregions 18 and 19 have a high radiation field which likely destroys the PAHs, while subregions 5 and 6 are toward the low-density edge of the \ion{H}{2} region where the gas might be optically thin to the ionizing flux and PAHs are destroyed.


The CCD in Figure \ref{N63_all}(d) shows a slight diagonal spread toward the top-left (most PAH). Subregions toward the center of N63 have by far the lowest fractional PAH emission and those farthest away have the highest. The fractional PAH emission appears to gradually increase radially without large jumps, likely because the PDRs are smeared by the large aperture size and line-of-sight integration for the subregions.

\texttt{DustEM} modeling shows that PAH fraction slightly increases outward from the central OB association of N63, consistent with the observations. ÊIn addition to its effect on PAH features, the OB association may also play a role in the increased $Y_{\rm VSG}$/$Y_{\rm BG}$ fraction (Figure \ref{N63_all}(b)), which peaks strongly at the locations of the highest radiation field and falls off along the radiation field.



\subsubsection{N180}
\begin{figure} Ê
\begin{center}
\includegraphics[scale=0.75]{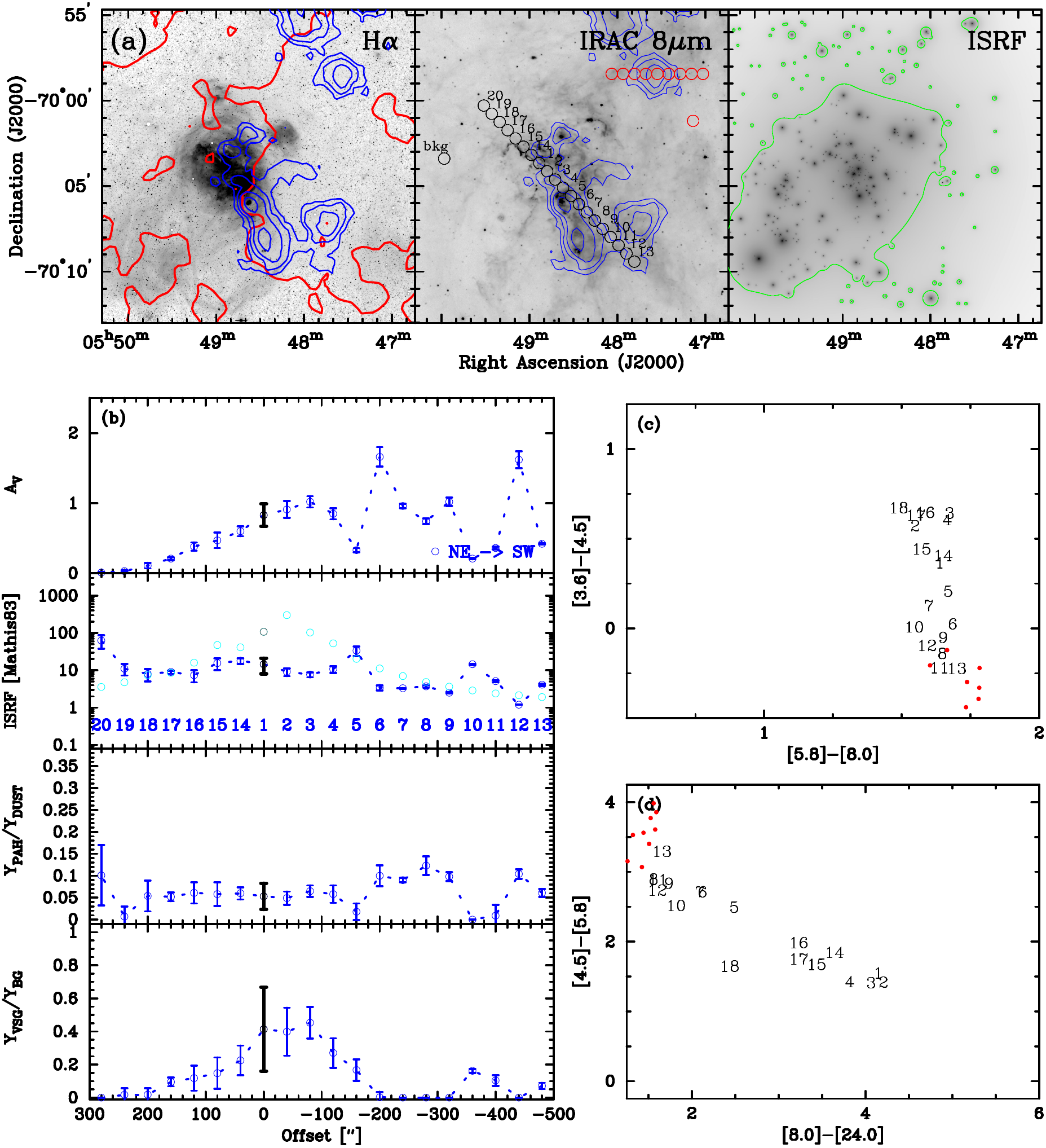}
\caption[Observations and Results of Classical \ion{H}{2}~Region N180]{Observations and results of classical \ion{H}{2} region N180. Description of plots are the same as Figure \ref{N63_all} with the exception of the blue contours showing large scale CO(1--0) emission from MAGMA \citep{Wong11}. The IRAC 8 $\mu$m image shows red circles at locations (along with a background subregion) where we test the prevalence of PAHs in molecular clouds in the CCD (also shown in red). \label{N180_all}
}
\end{center}
\end{figure}

N180 is a classical \ion{H}{2} region powered by a central OB-association, whose stars are concentrated at the center of the diagonal cut on subregions 1 through 3 (Figure \ref{N180_all}). ÊFollowing the diagonal cut southwest, molecular gas becomes prevalent in subregions 2 through 13. Northeast of the central OB association, the remaining subregions 14--20 contain little to no molecular gas and are located progressively further away from the massive stars. Ê



The CCD of N180 in Figure \ref{N180_all}(d)
exhibits three major subregion groups: (1) 1--4, (2) 14--17, and (3) 5--13, where subregions 19--20 are not significantly above the background and thus not included in the CCD. ÊGroup 1, located at the peaks of the ISRF and projected within the molecular cloud (Figure \ref{N180_all}(a)), has the highest [8.0]--[24.0] and the lowest [4.5]--[5.8] of all the subregion groups, i.e., it has the smallest PAH mass fraction.

Group 2 is located to the northeast of group 1 and is less affected by radiation from the massive stars. As the CCD indicates, PAH emission is increased in this group as compared to group 1, likely due to the fact that there is less harsh UV radiation to destroy PAHs. Moreover, PAH features are the most prevalent in group 3, whose subregions are deep in the molecular cloud and are far away from massive stars; thus, these locations have more available molecular gas and less destructive stellar radiation. The fitted results for the PAH fraction of dust agree with these results, with an overall increasing trend from the central subregions of group 1 toward group 3 and a dip in the PAH fraction for subregions 10 and 11 where there appears to be less molecular gas.


As with N63, the $Y_{\rm VSG}$/$Y_{\rm BG}$ fraction of N180 also peaks strongly at the locations of the highest radiation field (Figure \ref{N180_all}(b)). ÊFurthermore, it decreases rapidly to the southwest and more gradually to the northeast. ÊThe more dramatic decrease toward the southwest may be an indication of an increased fraction of BG inside the molecular cloud.

\subsection{Superbubbles--N70, N144}
\subsubsection{N70}
\begin{figure} Ê
\begin{center}
\includegraphics[scale=0.75]{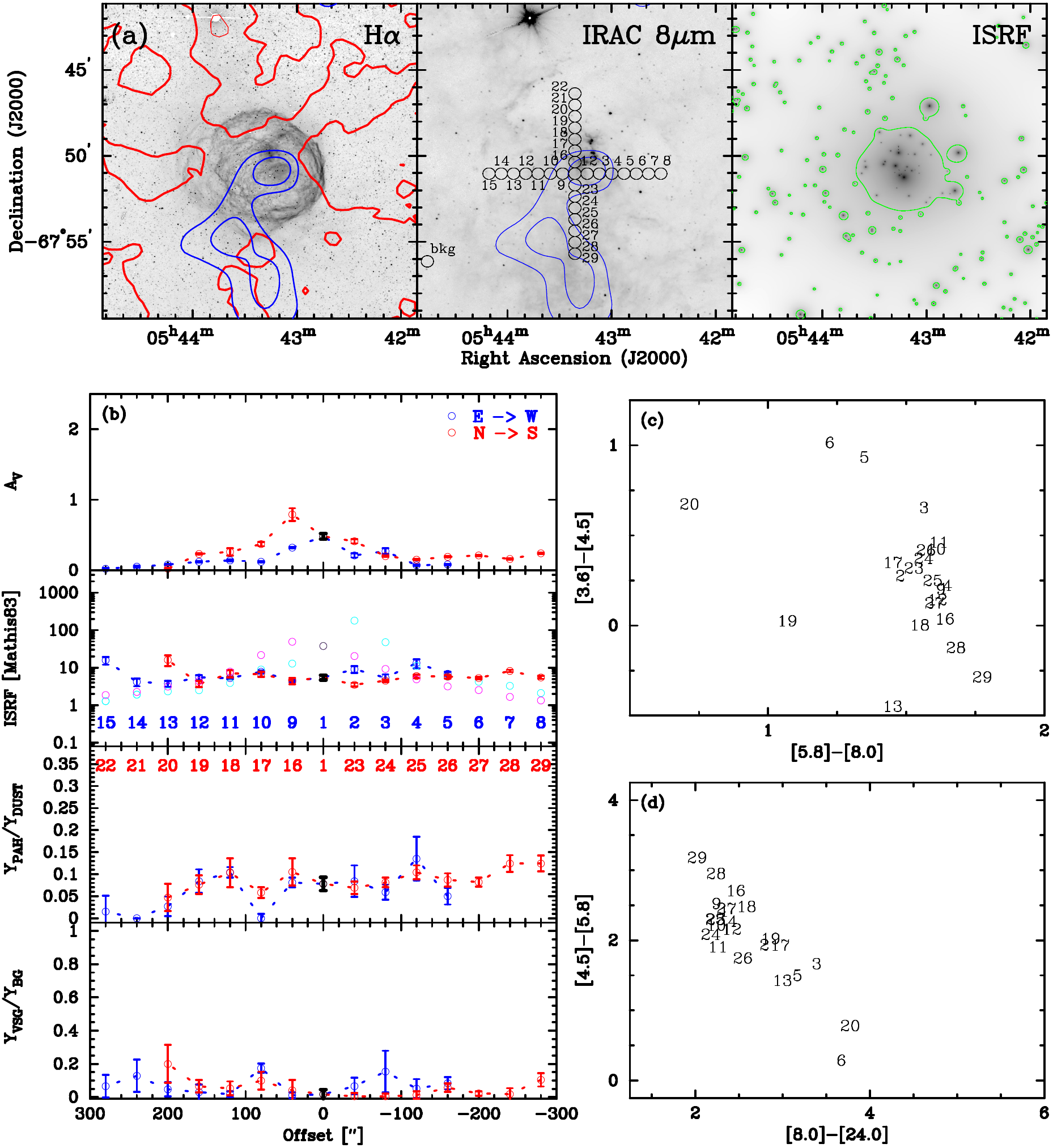}
\caption[Observations and Results of Supperbubble N70]{Observations and results of the superbubble N70. Description of plots are the same as Figure \ref{N63_all} with the exception of the blue contours showing large scale CO(1--0) emission from NANTEN \citep{Fukui08}.\label{N70_all}
}
\end{center}
\end{figure}

The N70 superbubble has a distinct shell and a hollow interior, with the central OB association dominating the ISRF particularly in subregions 1, 2, 3, 9, and 10 (Figure \ref{N70_all}). The SED flux is well-behaved compared to the other regions; fluxes generally decrease radially in each band, with the continuum bands decreasing slightly faster than PAH bands.

There are no obvious groupings in the CCDs (Figure \ref{N70_all}(c) and \ref{N70_all}(d)), though subregions 6 and 20 especially show signs of decreased PAH. These two subregions lie on the limb-brightened superbubble shell; at these locations there is increased emission at 4.5 $\mu$m (which is evident in Figure \ref{four_SB_fig2.ps}), possibly due to the Br$\alpha$ line incident in this band. Outside of the superbubble, the CCDs and \texttt{DustEM} fits show an increase in PAH emission for subregions 28 and 29, which are superimposed on CO and a small \ion{H}{1} concentration. The central molecular cloud, however, does not show an increase in PAH emission, but the radiation is high at this location.


While other regions have \texttt{DustEM} ISRF fits similar to the ISRF as determined by stars, the ISRF comparison is quite different in N70. This disparity may be due to the fact that stars are centralized in the superbubble, but the sampled subregions show all dust within the line of sight. As opposed to the other \ion{H}{2} regions, dust emission for N70 is likely coming from the shell and not the central region. Thus, a uniform distribution of dust along the hollow shell may explain why we do not find strong radial dependence on PAH or VSG. 



\subsubsection{N144}
\begin{figure} Ê
\begin{center}
\includegraphics[scale=0.75]{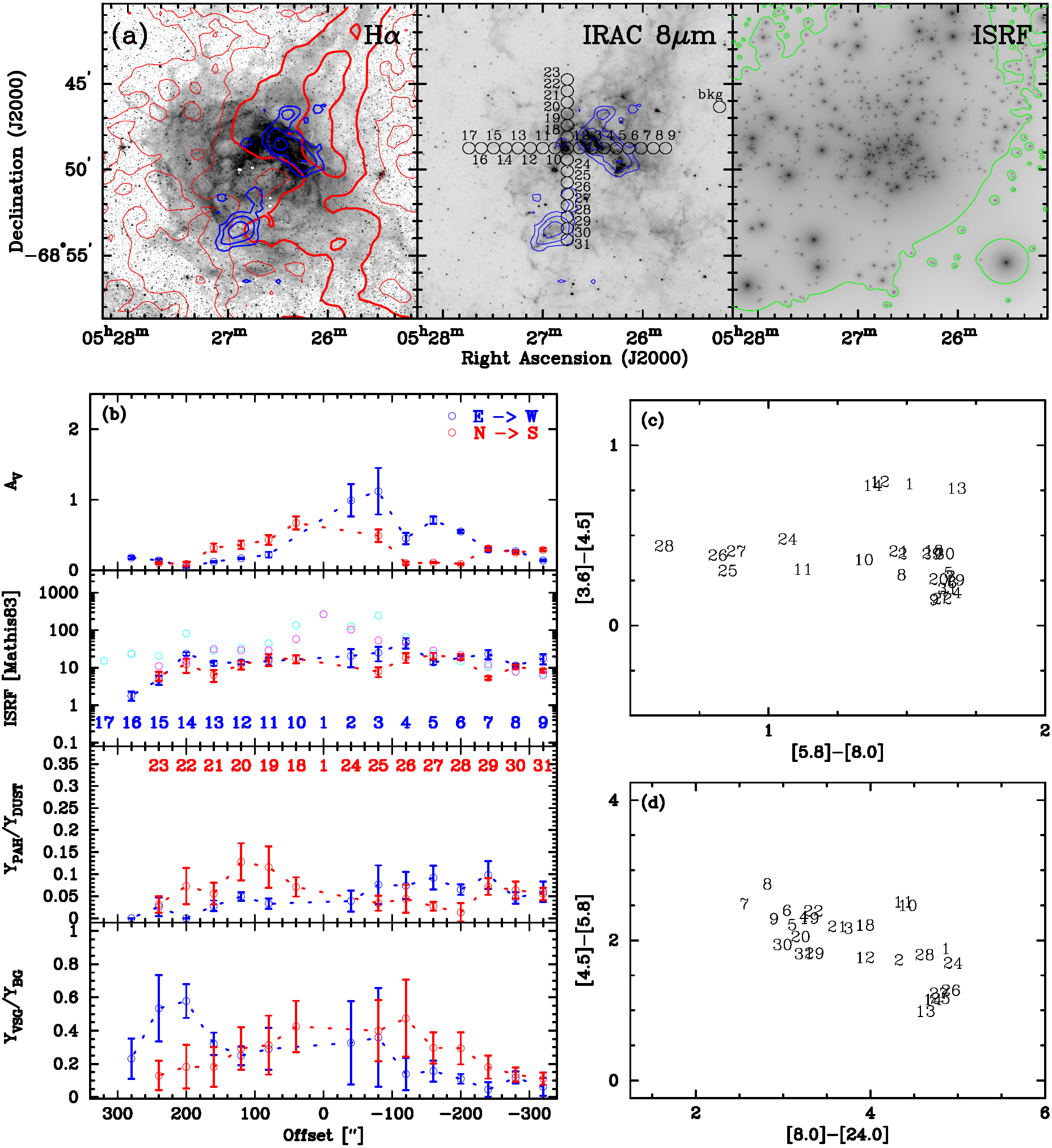}
\caption[Observations and Results of Supperbubble N144]{Observations and results of the supperbubble N144. Description of plots are the same as Figure \ref{N63_all} with the exception of the blue contours showing large scale CO(1--0) emission from MAGMA. \label{N144_all}
}
\end{center}
\end{figure}

N144 is a young supperbubble containing more massive stars than N70. Of the four regions, it is the only one declared optically thin by \citet{pel12}, although N70 is optically thin at most locations as well. ÊIndeed, N144 contains more massive stars (by number and mass) than any of our analyzed regions. Figure \ref{N144_all}(a) and (b) indicates a much higher radiation field, as compared to other regions, throughout all sampled subregions. Though massive stars are located throughout N144, their concentration peaks toward the center. Subregion~1 contains a massive YSO \citep{GC09} which dominates \emph{Spitzer} emission in all bands. Large-scale molecular clouds are found throughout N144, with the main clouds located toward the northwest and the south (Figure \ref{N144_all}(a)).

As indicated on the CCD in Figure \ref{N144_all}(d), subregions 13, 14, 25, 26, and 27 all have minimal fractional PAH emission. These subregions are located toward optically thin locations and, other than subregion 27, are outside molecular clouds. To the west, subregions 4--9 have the most fractional PAH emission, likely due to the presence of a molecular cloud (the 8 $\mu$m band indicates that subregions 7--9 are likely still in the molecular cloud) and a reduced radiation field. Subregions 2 and 3 show a fractional PAH emission transition toward that of subregions 4--9. In the south cut, subregions deepest in the molecular cloud (29--31) have the most fractional PAH emission, with subregion 28 showing a transition from subregions 25--27.

\texttt{DustEM} fits affirm a steady increase in PAH mass fraction in the south and west toward molecular clouds. Subregions 18--20 indicate the highest fitted PAH mass fractions, and we note that if we use lower contours in Figure \ref{N144_all}(a), there is large-scale molecular gas at these locations as well.

Fits of the $Y_{\rm{VSG}}/Y_{\rm{BG}}$ fraction are found to be the highest at the center, with a steady decrease toward the outer subregions. However, this decrease in $Y_{\rm{VSG}}/Y_{\rm{BG}}$ is not true for the eastern cut, where the fraction increases. The eastern cut is likely encountering an optically thin region (lacking [\ion{S}{2}]/H$\alpha$ enhancement), which may be responsible for an increase in the VSG population.

\subsection{Molecular Clouds and PAHs} \label{sec:mcp}
As seen in N180 and N144, the most obvious parameter that changes radially from the central OB association is the PAH mass fraction due to the presence of a (giant) molecular cloud. 
In order to further test if such features are genuine, we picked a molecular cloud to the northwest of N180 that is not associated with photoionized gas (i.e., no H$\alpha$ emission, Figure \ref{N180_all}(a)). The subregions in this molecular cloud are plotted as red points in the CCDs of N180 (Figure \ref{N180_all}(c) and (d)). These red points are in the same vicinity in the CCD as subregions 8--13 in N180, indicating a similar composition of PAH mass fraction in both molecular clouds.

Even on the outskirts of this northwest cloud, the red points stay in the same general vicinity in the CCD; however, N180 subregions 5--7 show transitory colors between the central OB association and the molecular cloud. For this northwest cloud, it is possible that PAHs are only excited on the surface of the cloud due to starlight and thus are not modified radially. N180, on the other hand, has coexistence of molecular gas and ionized gas, and intense radiation fields break down the PAHs. Thus, the destruction of PAHs by OB stars is likely to lead to changes in PAHs radially throughout an \ion{H}{2} region.

\section{Discussion} \label{sec:disc}
We have used model fits to spatially resolved SEDs of two classical and two superbubble \ion{H}{2} regions to determine the dust composition and properties with respect to the radiation field. In this section we discuss the individual dust components and compare the dust properties in differing \ion{H}{2} regions and SNRs.

\subsection{Dust}
\subsubsection{PAHs}

Radiation is needed to excite PAHs; however, PAHs also can be destroyed in the presence of a strong radiation field \citep[e.g.,][]{Madden06,Lebout07}, primarily due to high energy photons \citep[e.g.,][]{Aitken85,Voit92}. Moreover, the location of PAH and H$_2$ emission are strongly correlated in cool-PDRs, such as the filaments in the Horsehead Nebula \citep{Hab05,com07} and $\rho$ Ophiuchus \citep{Hab03}. However, the locations become anti-correlated with higher radiation field, as seen in the Orion bar \citep{Tie93} and in Monoceros R2 \citep{Ber09}.

The results in this paper are consistent with these studies. We find that PAH emission is enhanced toward locations of molecular clouds except where the radiation is high. We also find that PAH emission is diminished at locations that are optically thin to UV radiation.

\begin{figure}
\center
\includegraphics[scale=.45]{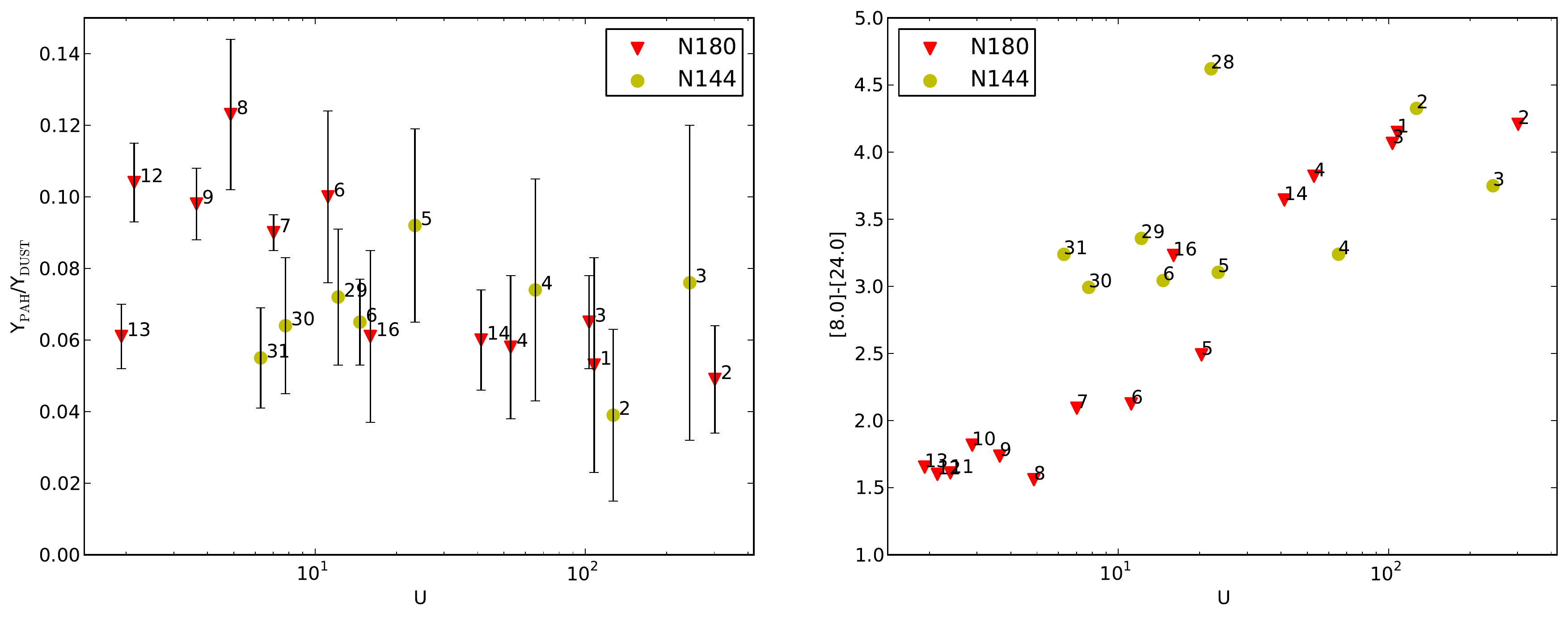}
\caption[$Y_{\rm PAH}$/$Y_{\rm DUST}$ versus Median Radiation Field]{Left panel: $Y_{\rm PAH}$/$Y_{\rm DUST}$ versus the median ISRF as calculated by massive stars. Points are only shown when the value of the fits are greater than the error bars. Right panel: The color [8.0]--[24.0] versus the median ISRF as calculated by massive stars. For both panels, subregions are shown only when at least half the region is a $\sim$2$\sigma$ detection for MAGMA CO ($\sigma$ = 0.3 K km s$^{-1}$). The number next to each point shows the corresponding subregion.
\label{fig:Upah}}
\end{figure}

Since our regions are relatively young (no known SNRs in N144 and N180, one known in N63, and a history of one or multiple in N70), the radiation field is dominated by massive stars that produce harsh photons. Although high energy photons are the main mechanism for destroying PAHs rather than the radiation field, we will assume that the stellar content of the regions follow the same initial mass function and have similar spectral shape in their radiation field and attempt to provide an estimate of the typical $U$ required for significant PAH destruction. In particular, we focus on N144 and N180 since N63 lacks large-scale CO observations and the dust emission from N70 is thought to come from the surrounding shell. For these two regions, Figure \ref{fig:Upah} both $Y_{\rm{PAH}}/Y_{\rm{DUST}}$ and the color [8.0]--[24.0] versus the radiation field as ascertained by massive stars ($U$). We only show subregions when at least half the region contains a 2$\sigma$ Magellanic Mopra Assessment \citep[MAGMA;][]{Wong11} CO detection. Note that higher values of the color [8.0]--[24.0] indicates less PAH emission. For both panels, subregions for N180 indicate that PAHs are less coincident with CO molecular gas for high $U$, particularly above $U \approx 30$. N144 subregions have points in the same general vicinity as N180, but an obvious correlation between PAHs and radiation field is not apparent. This could be due to the lack of data (particularly at low values of $U$) or physical differences between N144 and N180. 

Our calculations of the radiation field may be highly uncertain since they can be quite dissimilar (sometimes by about an order of magnitude) to the fitted results. However, the fitted results of $U$ are typically below those of the massive stars, so $U \approx 30$ is a fair estimate of an upper limit of the typical radiation field needed to destroy PAHs in N180. Though the $U$ limit of N180 may not be representative of other molecular clouds, we note that maps in \citet{Galametz13} also show a reduction at PAHs with similar modeled $U$ values in the LMC classical \ion{H}{2} region N159, with a drastic depletion in the fraction of PAHs for $U \gtrsim$ 100.

\subsubsection{VSGs and BGs} \label{sec:VSGBG}
\begin{figure}
\center
\includegraphics[scale=.45]{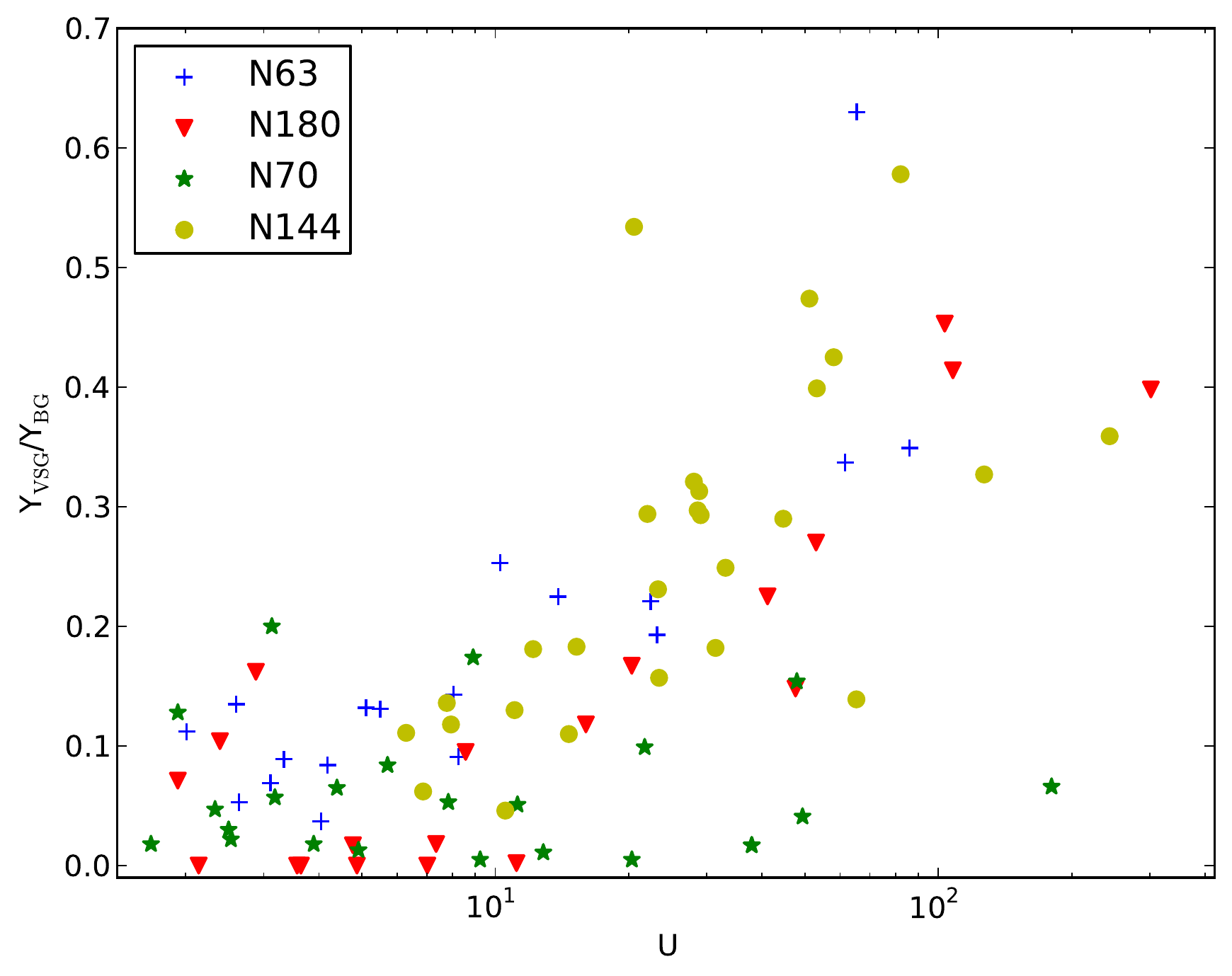}
\caption[$Y_{\rm VSG}$/$Y_{\rm BG}$ versus Median Radiation Field]{$Y_{\rm VSG}$/$Y_{\rm BG}$ versus the median ISRF as calculated by massive stars for all four regions. Error bars for $Y_{\rm VSG}$/$Y_{\rm BG}$ are not shown for clarity. \label{fig:U_vs_vsg_bg_ratio-eps-converted-to.pdf}}
\end{figure}

In all of our \ion{H}{2} regions except N70 (whose dust is uniformly distributed about its shell), $Y_{\rm VSG}$/$Y_{\rm BG}$ is highest at locations where the radiation field is the strongest, as evident in Figure \ref{fig:U_vs_vsg_bg_ratio-eps-converted-to.pdf}. We stress that the VSG population is highly dependent on each SED's 24~$\mu$m excess and is typically anti-correlated with the BG population.

Grain populations typically change due to both destruction (i.e., dust changes back to gas) or disruption (large grains break into smaller grains). \citet{Jones96} showed that fast shocks ($>$150 km~s$^{-1}$) likely destroy grains via sputtering, while the velocity requirement for shattering BGs with VSGs depends on the dust species, with velocities ranging from $\sim$1--3 km~s$^{-1}$. In our regions, dust-dust collisions are probably uncommon due to low densities ($n_{\rm{H}} \sim 100$~cm$^{-3}$). However, O-stars are known to have wind-blown bubbles expanding at velocities of 10--15 km~s$^{-1}$ \citep[with observations of such bubbles in N180,][]{Naze01}, which could lead to grain shattering.

Other possibilities of an increased $Y_{\rm VSG}$ include dust coagulation and dust erosion. Dust coagulation of the smallest grains into slightly larger VSGs could increase the $24~\mu$m emission, but the dust in our regions are not in environments conducive to coagulation \citep[see, e.g.,][]{koh12}. \citet{Betal08} suggested erosion of BGs into VSGs likely explains the LMC's 70~$\mu$m excess in neutral/ionized regions. Moreover, dust erosion is thought to be increased during star formation due to increased dust processing \citep{par11}. The central OB associations of our \ion{H}{2} regions have a rich history of previous star formation, which could certainly lead to more dust processing/erosion than the areas analyzed at the macro-scale (several 100~pc; \citealt{Betal08}), leading to the increased 24~$\mu$m emission. Thus, at scales of $\sim$10~pc, BGs near OB associations may have gone through substantial erosion that has caused this increase of VSGs.


On the other hand, \citet{Pilleri2012} saw a definite decrease of ``evaporating" VSGs with increasing radiation field. The sources they analyzed typically had a higher radiation field ($U \gtrsim 100$) than our regions, though we still see an increase in VSGs from $U=100$ to $U=1000$. However, their dust model of VSGs follows more closely to the smallest grains of \citet{Li01} while ours follows \citet{Detal90}. Since \citet{Pilleri2012} modeled VSGs that are more similar to PAHs and thus have an emission spectrum at much shorter wavelengths, these VSGs are not comparable to the VSGs modeled in this paper.

\subsection{Dust Properties of Evolved \ion{H}{2}~Regions}
We find that N70, the most evolved \ion{H}{2} region, likely has its dust mass located in the shell. However, major differences in dust properties between classical \ion{H}{2} regions and superbubbles are not obvious with this small sample size. \citet{Relano13} analyzes the SEDs of 119 \ion{H}{2} regions in M33, categorizing them in a likely evolutionary sequence (filled, mixed, shell, and clear shell morphologies). They found that each category had similar SED features, though differed slightly between each classification. Specifically, they found that the FIR peak is located at longer wavelengths for older regions (shells and clear shells), indicating that colder dust is likely present. Uniform cold dust temperatures are expected since gas has been pushed far away from the central heating source. This result is consistent with our analysis of N70, which shows a uniform, low fraction of $Y_{\rm{VSG}}/Y_{\rm{BG}}$ compared to our less evolved \ion{H}{2} regions that do not have the same shell geometry.

\subsection{Comparison to SNR Studies}
Changes in dust populations are commonly studied in SNRs due to their dynamic properties. Fast shocks ($\gtrsim150$ km~s$^{-1}$) allow dust to reach high velocities that may cause their destruction via catastrophic shattering or sputtering \citep[e.g.,][]{Jones96,Micelotta10}, though VSGs and PAHs can also be created through low-velocity shattering \citep{Jones96}. While PAHs can survive or even be produced in slow shocks, fast shocks are generally considered to destroy all PAHs in a region except in circumstances where they can be recreated or protected by their environment \citep[e.g.,][]{Micelotta10,Seok12}. Additionally, harsh photons provide another mechanism of destroying the smallest grains; the radiation field in SNRs can vary drastically, though it is typically about 10-100 times the local ISRF \citep[e.g.,][]{Andersen11}. With these different mechanisms for creation and destruction of dust, it is important to observe and model a wide range of SNRs with different shock velocities, densities, and radiation fields to test theoretical models. Moreover, dust properties of SNRs can be compared to our \ion{H}{2} regions to provide a diagnostic of how dust changes in varying radiation fields in the absence of large shock velocities.


The \emph{Spitzer} IR Spectrograph (IRS) has been used to study a large number of SNRs.  Dust produced in supernova ejecta has been detected in young ($<$300 yr) SNRs \citep{Rho09}, and PAHs are detected in SNRs \citep[e.g.,][]{Tappe06}.  Using a large spectral survey of 14 Galactic SNRs, \citet{Andersen11} used \texttt{DustEM} (also using the \citealt{Detal90} dust models) to model the contributions of emission from BGs, VSGs, and PAHs. They found that the ratio of VSGs to BGs is 2--3 times larger in SNRs than the plane of the Galaxy, and they attribute dust shattering as the mechanism for creating the surplus of VSGs. Two of the 14 SNRs had relatively high radiation fields; fits of these regions had relatively low PAHs and virtually no VSGs. \citet{Andersen11} suggested that these regions have high shock velocities and low densities, which can lead to destruction of VSGs and PAHs through sputtering.

With high radiation, we find that the VSG population can increase and PAHs are generally destroyed; however, fast shocks can destroy both grains. Unlike SNR regions, we find that VSG population increases where there is higher $U$, and there are no known fast shocks in our regions. Indeed, while observations from \emph{XMM Newton} show X-ray emission from N70, there is no X-ray emission toward N180 and N144, and for N63, X-ray emission is only associated with the known SNR N63A (S. L. Snowden et al. in preparation). It has been suggested that N70 has recently been heated by an interior SNR \citep{CM90,rod11,zha14}. If shocks are the main mechanism that are heating the grains in N70 rather than the ISRF, our \texttt{DustEM} modeling would be invalid (since $U$ is our free parameter rather than temperature). Regardless, the shape of the SEDs of each subregion is the same with a dip at 24~$\mu$m, which agree with our \texttt{DustEM} fit of a uniform, low abundance of VSG. Therefore, it is most likely that a past history of fast shocks has destroyed the population of VSGs, and no mechanism has been available for creation of new VSGs (due to, e.g., the lack of centrally located gas since it has been blown away). 

\section{Summary} Ê\label{sec:sum}
This paper is the first to examine radial SEDs on small scales ($\sim$10~pc) across \ion{H}{2} regions (of size-scales of 100-200~pc) in the LMC in order to evaluate the changes of dust properties caused by energy feedback of massive stars

We selected two classical \ion{H}{2} regions (N63 and N180) and two superbubbles (N70 and N144) in the LMC. From the massive stellar population, we approximated the radiation field throughout each region. ÊSEDs were constructed using the seven \emph{Spitzer} bands from the LMC (four IRAC and three MIPS bands). We used CCDs and \texttt{DustEM} modeling to analyze the changes in SEDs radially throughout each \ion{H}{2} region. The changes in the SEDs were then compared to the radiation field in order to gauge the effects of massive stars on dust properties.

Our main results are as follows:

1) We find that the PAH mass fraction increases significantly toward molecular clouds except when there is a very strong radiation field. PAHs are likely being destroyed by the radiation field (in areas with $U\gtrsim$ 30 for N180), and we typically detect the PAH mass fraction increase as we leave the central OB association. As expected, optically thin areas in our \ion{H}{2} regions also have a diminished PAH mass fraction.

2) The VSG mass fraction increases at locations of an enhanced radiation field. Expanding bubbles may be launching dust at velocities that can cause BGs to shatter into VSGs. Dust erosion of larger grains into VSGs is another possible mechanism for the enhanced 24~$\mu$m emission.



\acknowledgments
This research was supported by NASA grant HST-GO-12941 06-A. This research has also made use of the NASA/IPAC Infrared
Science Archive, which is operated by the Jet Propulsion
Laboratory, California Institute of Technology, under contract
with the National Aeronautics and Space Administration.

\newpage
\begin{deluxetable}{ccccccccc}
\tabletypesize{\scriptsize}
\tablewidth{0pc}
\tablecaption{Photometric Flux Density Measurements\tablenotemark{a}\label{tab_all_fl}}
\tablehead{
\colhead{HII}	&
\colhead{Subregion}    	&
\colhead{3.6\um}      	&
\colhead{4.5\um}     	&
\colhead{5.8\um}      	&
\colhead{8.0\um}     	&
\colhead{24\um}     	&
\colhead{70\um}      	&
\colhead{160\um}     	\\
\colhead{Region}            	&
\colhead{}            	&  
\colhead{(mJy)}      	&
\colhead{(mJy)}     	&
\colhead{(mJy)}      	&
\colhead{(mJy)}     	&
\colhead{(mJy)}     	&
\colhead{(mJy)}      	&
\colhead{(mJy)}   	 
}
\startdata
N63& 1 & $13.4\,^{+31.1}_{-7.9}$ & $12.9\,^{+29.4}_{-7.2}$ & $48.1\,^{+145.6}_{-27.0}$ & $119.1\,^{+316.3}_{-75.6}$ & $933\,^{+804.9}_{-406.0}$ & $8513\,^{+6715.9}_{-3565.4}$ & $3186\,^{+436.8}_{-637.3}$ \\
& 2 & $3.4\,^{+1.2}_{-0.6}$ & $3.5\,^{+0.9}_{-0.5}$ & $10.7\,^{+2.4}_{-1.7}$ & $22.5\,^{+4.7}_{-3.7}$ & $121\,^{+24.7}_{-25.9}$ & $1372\,^{+351.8}_{-306.9}$ & $1956\,^{+52.3}_{-87.3}$ \\
& 3 & $2.2\,^{+0.6}_{-0.5}$ & $2.3\,^{+0.3}_{-0.3}$ & $8.4\,^{+2.9}_{-1.5}$ & $18.3\,^{+7.5}_{-4.1}$ & $72\,^{+9.4}_{-5.3}$ & $920\,^{+70.1}_{-48.3}$ & $1397\,^{+40.0}_{-42.6}$ \\
& 4 & $2.3\,^{+0.8}_{-0.6}$ & $2.4\,^{+0.5}_{-0.4}$ & $8.2\,^{+3.8}_{-1.8}$ & $19.1\,^{+9.8}_{-5.9}$ & $51\,^{+10.5}_{-6.2}$ & $702\,^{+39.1}_{-67.3}$ & $1014\,^{+6.3}_{-18.0}$ \\
& 5 & $0.8\,^{+0.3}_{-0.2}$ & $1.2\,^{+0.2}_{-0.2}$ & $3.7\,^{+1.1}_{-1.1}$ & $7.5\,^{+1.4}_{-1.3}$ & $21\,^{+3.4}_{-2.4}$ & $426\,^{+39.0}_{-15.9}$ & $635\,^{+15.5}_{-7.5}$ \\
& 6 & $0.9\,^{+0.7}_{-0.5}$ & $0.9\,^{+0.4}_{-0.4}$ & $3.7\,^{+1.5}_{-1.0}$ & $6.1\,^{+1.6}_{-1.4}$ & $13\,^{+1.6}_{-1.8}$ & $283\,^{+20.2}_{-20.4}$ & $564\,^{+38.4}_{-26.2}$ \\
& 7 & $0.4\,^{+0.3}_{-0.3}$ & $0.3\,^{+0.3}_{-0.2}$ & $1.7\,^{+0.9}_{-0.7}$ & $1.6\,^{+2.5}_{-1.4}$ & $8\,^{+2.9}_{-2.2}$ & $322\,^{+72.0}_{-60.9}$ & $740\,^{+43.4}_{-167.8}$ \\
& 8 & $3.2\,^{+0.7}_{-0.6}$ & $3.4\,^{+0.6}_{-0.5}$ & $13.9\,^{+3.5}_{-2.4}$ & $31.5\,^{+7.2}_{-5.5}$ & $481\,^{+46.9}_{-42.8}$ & $2704\,^{+589.7}_{-407.9}$ & $3040\,^{+394.7}_{-413.9}$ \\
& 9 & $3.4\,^{+5.2}_{-1.5}$ & $2.4\,^{+4.2}_{-1.0}$ & $16.7\,^{+21.0}_{-5.9}$ & $42.2\,^{+53.3}_{-15.3}$ & $70\,^{+67.0}_{-31.4}$ & $1481\,^{+692.0}_{-619.2}$ & $2718\,^{+104.8}_{-274.0}$ \\
&10 & $2.0\,^{+3.4}_{-1.6}$ & $1.2\,^{+1.6}_{-1.0}$ & $8.1\,^{+16.8}_{-5.1}$ & $16.5\,^{+47.5}_{-13.5}$ & $19\,^{+20.4}_{-12.5}$ & $182\,^{+125.3}_{-90.7}$ & $1171\,^{+331.7}_{-407.6}$ \\
&11 & $0.3\,^{+0.5}_{-0.2}$ & $0.4\,^{+0.4}_{-0.3}$ & $1.0\,^{+1.0}_{-0.9}$ & $-0.6\,^{+1.3}_{-1.0}$ & $2\,^{+1.1}_{-1.1}$ & $-15\,^{+23.7}_{-11.4}$ & $197\,^{+39.5}_{-64.5}$ \\
&12 & $-0.0\,^{+0.3}_{-0.2}$ & $0.0\,^{+0.3}_{-0.3}$ & $-0.1\,^{+1.0}_{-0.9}$ & $-2.6\,^{+0.7}_{-0.6}$ & $0\,^{+0.9}_{-0.8}$ & $-60\,^{+18.2}_{-16.3}$ & $3\,^{+11.2}_{-29.1}$ \\
&13 & $4.7\,^{+1.1}_{-0.8}$ & $5.3\,^{+1.2}_{-0.7}$ & $16.2\,^{+3.5}_{-2.5}$ & $36.7\,^{+8.5}_{-6.8}$ & $253\,^{+51.7}_{-36.9}$ & $2301\,^{+79.9}_{-425.0}$ & $2505\,^{+43.7}_{-93.9}$ \\
&14 & $3.2\,^{+1.2}_{-1.0}$ & $3.2\,^{+0.9}_{-0.7}$ & $14.3\,^{+4.9}_{-3.2}$ & $32.2\,^{+12.1}_{-7.9}$ & $97\,^{+21.4}_{-14.5}$ & $1286\,^{+135.0}_{-71.2}$ & $2245\,^{+50.3}_{-60.9}$ \\
&15 & $4.3\,^{+1.6}_{-1.2}$ & $3.3\,^{+0.9}_{-0.8}$ & $21.7\,^{+6.3}_{-4.9}$ & $53.0\,^{+18.2}_{-11.1}$ & $77\,^{+9.7}_{-10.4}$ & $1045\,^{+76.2}_{-80.4}$ & $2016\,^{+78.2}_{-263.2}$ \\
&16 & $2.3\,^{+0.6}_{-0.5}$ & $2.3\,^{+0.6}_{-0.4}$ & $11.8\,^{+2.5}_{-2.2}$ & $26.6\,^{+7.3}_{-3.9}$ & $47\,^{+7.9}_{-6.8}$ & $1035\,^{+63.4}_{-69.2}$ & $1550\,^{+9.3}_{-181.5}$ \\
&17 & $2.2\,^{+1.1}_{-0.9}$ & $2.1\,^{+1.1}_{-0.9}$ & $9.8\,^{+3.9}_{-3.3}$ & $21.6\,^{+10.6}_{-9.1}$ & $30\,^{+16.6}_{-9.5}$ & $849\,^{+161.5}_{-185.1}$ & $828\,^{+17.7}_{-22.0}$ \\
&18 & $1.4\,^{+0.7}_{-0.4}$ & $1.9\,^{+0.7}_{-0.4}$ & $4.8\,^{+1.7}_{-1.7}$ & $7.9\,^{+2.5}_{-2.5}$ & $130\,^{+101.0}_{-41.2}$ & $1442\,^{+882.0}_{-529.3}$ & $1401\,^{+291.7}_{-159.8}$ \\
&19 & $1.4\,^{+0.4}_{-0.3}$ & $1.7\,^{+0.4}_{-0.3}$ & $4.4\,^{+1.8}_{-1.5}$ & $8.3\,^{+4.6}_{-2.5}$ & $44\,^{+5.8}_{-6.7}$ & $572\,^{+35.5}_{-11.8}$ & $1141\,^{+20.9}_{-32.3}$ \\
&20 & $1.7\,^{+0.4}_{-0.4}$ & $1.7\,^{+0.4}_{-0.4}$ & $6.8\,^{+1.3}_{-1.6}$ & $16.6\,^{+3.4}_{-2.7}$ & $36\,^{+2.3}_{-3.0}$ & $596\,^{+55.8}_{-48.0}$ & $1116\,^{+45.6}_{-18.7}$ \\
&21 & $1.2\,^{+0.4}_{-0.3}$ & $1.1\,^{+0.3}_{-0.2}$ & $5.8\,^{+1.6}_{-1.3}$ & $13.4\,^{+3.9}_{-2.1}$ & $23\,^{+2.2}_{-2.2}$ & $448\,^{+10.9}_{-15.0}$ & $1171\,^{+31.1}_{-44.3}$ \\
&22 & $1.9\,^{+0.4}_{-0.4}$ & $1.5\,^{+0.3}_{-0.3}$ & $10.7\,^{+2.0}_{-2.6}$ & $28.1\,^{+5.2}_{-6.3}$ & $32\,^{+3.2}_{-5.2}$ & $471\,^{+41.6}_{-37.8}$ & $1229\,^{+49.1}_{-4.5}$ \\
\hline
N180& 1 & $15.2\,^{+5.4}_{-3.9}$ & $13.9\,^{+3.2}_{-2.1}$ & $38.6\,^{+20.3}_{-9.9}$ & $97.3\,^{+51.9}_{-31.1}$ & $495\,^{+113.3}_{-59.4}$ & $6103\,^{+1642.5}_{-1140.2}$ & $6340\,^{+55.4}_{-636.6}$ \\
& 2 & $9.7\,^{+2.1}_{-1.3}$ & $10.7\,^{+1.6}_{-1.1}$ & $26.3\,^{+4.8}_{-4.4}$ & $61.2\,^{+11.9}_{-10.2}$ & $329\,^{+40.7}_{-26.4}$ & $3952\,^{+499.4}_{-386.6}$ & $4919\,^{+795.3}_{-237.4}$ \\
& 3 & $10.2\,^{+1.5}_{-1.4}$ & $12.1\,^{+1.1}_{-0.9}$ & $29.5\,^{+7.9}_{-5.4}$ & $76.8\,^{+19.9}_{-16.4}$ & $362\,^{+35.1}_{-24.0}$ & $3569\,^{+162.7}_{-322.4}$ & $4994\,^{+52.4}_{-454.9}$ \\
& 4 & $10.8\,^{+2.2}_{-1.7}$ & $12.3\,^{+1.6}_{-1.0}$ & $30.5\,^{+20.5}_{-6.0}$ & $78.9\,^{+55.8}_{-17.8}$ & $297\,^{+31.8}_{-41.3}$ & $3962\,^{+451.1}_{-500.3}$ & $5154\,^{+443.8}_{-102.1}$ \\
& 5 & $8.6\,^{+4.4}_{-3.0}$ & $6.8\,^{+3.2}_{-1.6}$ & $45.0\,^{+21.3}_{-13.8}$ & $116.8\,^{+55.3}_{-39.3}$ & $129\,^{+53.3}_{-35.2}$ & $2504\,^{+900.2}_{-544.9}$ & $4364\,^{+97.1}_{-13.6}$ \\
& 6 & $6.1\,^{+2.3}_{-1.9}$ & $4.1\,^{+1.2}_{-1.1}$ & $33.1\,^{+9.2}_{-10.8}$ & $87.3\,^{+24.4}_{-27.9}$ & $69\,^{+12.1}_{-20.0}$ & $1299\,^{+198.7}_{-273.6}$ & $4271\,^{+133.6}_{-712.3}$ \\
& 7 & $3.0\,^{+0.8}_{-0.6}$ & $2.2\,^{+0.5}_{-0.4}$ & $17.9\,^{+1.6}_{-1.3}$ & $43.6\,^{+4.3}_{-3.2}$ & $33\,^{+1.8}_{-2.2}$ & $717\,^{+26.1}_{-32.6}$ & $2418\,^{+138.9}_{-31.3}$ \\
& 8 & $5.0\,^{+1.3}_{-1.2}$ & $2.9\,^{+0.8}_{-0.7}$ & $27.5\,^{+5.7}_{-6.2}$ & $69.9\,^{+17.6}_{-17.2}$ & $32\,^{+5.9}_{-7.5}$ & $642\,^{+70.1}_{-95.2}$ & $1984\,^{+229.1}_{-37.5}$ \\
& 9 & $3.2\,^{+1.2}_{-0.7}$ & $2.0\,^{+0.7}_{-0.5}$ & $18.1\,^{+3.0}_{-2.5}$ & $46.3\,^{+5.6}_{-5.7}$ & $25\,^{+2.5}_{-2.6}$ & $484\,^{+22.7}_{-23.1}$ & $2037\,^{+86.0}_{-165.4}$ \\
&10 & $3.3\,^{+2.2}_{-1.0}$ & $2.2\,^{+1.5}_{-0.7}$ & $14.8\,^{+2.5}_{-2.4}$ & $34.2\,^{+6.5}_{-5.8}$ & $20\,^{+3.3}_{-2.7}$ & $327\,^{+37.6}_{-19.6}$ & $2182\,^{+0.7}_{-214.1}$ \\
&11 & $4.0\,^{+1.2}_{-0.8}$ & $2.1\,^{+0.8}_{-0.5}$ & $20.2\,^{+4.0}_{-2.9}$ & $50.9\,^{+9.7}_{-7.7}$ & $25\,^{+1.7}_{-2.2}$ & $324\,^{+25.3}_{-25.2}$ & $1750\,^{+98.0}_{-4.6}$ \\
&12 & $3.6\,^{+1.8}_{-1.0}$ & $2.1\,^{+1.1}_{-0.7}$ & $17.6\,^{+2.0}_{-2.3}$ & $42.6\,^{+3.9}_{-4.2}$ & $20\,^{+1.6}_{-1.5}$ & $243\,^{+25.3}_{-25.2}$ & $1794\,^{+32.6}_{-605.0}$ \\
&13 & $2.1\,^{+0.8}_{-0.4}$ & $1.1\,^{+0.5}_{-0.3}$ & $15.6\,^{+1.8}_{-1.5}$ & $41.8\,^{+3.5}_{-2.3}$ & $21\,^{+3.6}_{-1.8}$ & $241\,^{+22.9}_{-43.1}$ & $1377\,^{+32.7}_{-4.6}$ \\
&14 & $11.2\,^{+2.6}_{-2.0}$ & $10.6\,^{+2.0}_{-1.6}$ & $38.6\,^{+8.3}_{-4.5}$ & $98.5\,^{+21.2}_{-10.0}$ & $315\,^{+79.6}_{-48.6}$ & $5091\,^{+713.4}_{-285.3}$ & $5127\,^{+275.8}_{-237.4}$ \\
&15 & $8.6\,^{+4.9}_{-1.9}$ & $8.4\,^{+2.6}_{-1.4}$ & $26.1\,^{+9.5}_{-5.0}$ & $61.9\,^{+22.8}_{-10.9}$ & $163\,^{+32.9}_{-26.1}$ & $3201\,^{+576.4}_{-353.7}$ & $3542\,^{+908.0}_{-518.0}$ \\
&16 & $2.2\,^{+0.8}_{-0.5}$ & $2.6\,^{+0.6}_{-0.5}$ & $10.7\,^{+2.0}_{-2.2}$ & $25.6\,^{+3.8}_{-5.4}$ & $56\,^{+13.6}_{-10.3}$ & $1043\,^{+327.4}_{-229.3}$ & $1732\,^{+167.7}_{-306.7}$ \\
&17 & $1.6\,^{+0.6}_{-0.4}$ & $1.8\,^{+0.4}_{-0.3}$ & $6.1\,^{+1.4}_{-1.1}$ & $14.2\,^{+4.0}_{-2.6}$ & $31\,^{+2.5}_{-4.1}$ & $702\,^{+26.9}_{-54.0}$ & $1078\,^{+48.5}_{-88.8}$ \\
&18 & $0.9\,^{+0.8}_{-0.4}$ & $1.1\,^{+0.5}_{-0.4}$ & $3.3\,^{+1.9}_{-1.4}$ & $7.2\,^{+2.9}_{-3.6}$ & $7\,^{+2.3}_{-2.8}$ & $294\,^{+103.0}_{-11.9}$ & $528\,^{+194.2}_{-176.1}$ \\
&19 & $0.2\,^{+0.5}_{-0.3}$ & $0.3\,^{+0.4}_{-0.3}$ & $0.8\,^{+0.8}_{-0.9}$ & $0.5\,^{+1.4}_{-1.4}$ & $1\,^{+1.0}_{-1.0}$ & $118\,^{+37.9}_{-30.5}$ & $175\,^{+10.7}_{-14.9}$ \\
&20 & $-0.1\,^{+0.3}_{-0.3}$ & $0.0\,^{+0.3}_{-0.3}$ & $1.9\,^{+0.8}_{-1.1}$ & $3.8\,^{+1.6}_{-1.9}$ & $3\,^{+1.6}_{-1.3}$ & $99\,^{+6.9}_{-11.0}$ & $55\,^{+131.3}_{-1.1}$ \\
\hline
N70& 1 & $2.6\,^{+0.9}_{-0.4}$ & $2.0\,^{+0.6}_{-0.3}$ & $12.4\,^{+2.9}_{-1.3}$ & $31.1\,^{+7.4}_{-2.5}$ & $29\,^{+11.3}_{-5.0}$ & $766\,^{+122.0}_{-102.2}$ & $1744\,^{+78.0}_{-251.5}$ \\
& 2 & $3.2\,^{+1.6}_{-1.1}$ & $2.7\,^{+0.9}_{-0.6}$ & $10.5\,^{+7.0}_{-3.8}$ & $22.9\,^{+18.8}_{-8.5}$ & $33\,^{+3.7}_{-4.5}$ & $696\,^{+106.3}_{-88.2}$ & $1088\,^{+174.6}_{-81.3}$ \\
& 3 & $1.5\,^{+0.6}_{-0.3}$ & $1.8\,^{+0.4}_{-0.5}$ & $5.5\,^{+1.0}_{-0.9}$ & $13.0\,^{+1.8}_{-2.2}$ & $33\,^{+21.5}_{-8.9}$ & $519\,^{+88.3}_{-74.7}$ & $983\,^{+41.6}_{-158.8}$ \\
& 4 & $1.8\,^{+1.2}_{-0.5}$ & $1.5\,^{+0.6}_{-0.4}$ & $8.0\,^{+4.1}_{-1.6}$ & $20.4\,^{+9.6}_{-4.1}$ & $21\,^{+6.3}_{-1.9}$ & $365\,^{+51.1}_{-72.9}$ & $464\,^{+113.7}_{-24.7}$ \\
& 5 & $0.5\,^{+0.5}_{-0.3}$ & $0.8\,^{+0.4}_{-0.3}$ & $2.1\,^{+1.0}_{-0.7}$ & $4.1\,^{+2.5}_{-1.4}$ & $8\,^{+1.3}_{-1.4}$ & $191\,^{+15.2}_{-18.0}$ & $345\,^{+9.6}_{-29.0}$ \\
& 6 & $1.1\,^{+1.0}_{-0.8}$ & $1.9\,^{+1.4}_{-1.0}$ & $1.7\,^{+1.4}_{-1.2}$ & $2.9\,^{+1.8}_{-2.3}$ & $9\,^{+2.1}_{-4.5}$ & $196\,^{+21.9}_{-56.5}$ & $27\,^{+25.1}_{-25.7}$ \\
& 7 & $0.1\,^{+0.4}_{-0.2}$ & $0.1\,^{+0.3}_{-0.2}$ & $-0.3\,^{+0.9}_{-0.9}$ & $-1.2\,^{+1.1}_{-0.8}$ & $0\,^{+1.4}_{-1.3}$ & $-15\,^{+30.4}_{-16.9}$ & $-103\,^{+27.4}_{-0.3}$ \\
& 8 & $0.3\,^{+0.5}_{-0.3}$ & $0.2\,^{+0.4}_{-0.3}$ & $-0.8\,^{+1.0}_{-0.9}$ & $-1.7\,^{+0.9}_{-0.9}$ & $0\,^{+1.1}_{-1.2}$ & $-31\,^{+20.0}_{-11.3}$ & $-109\,^{+22.7}_{-1.2}$ \\
& 9 & $1.3\,^{+0.3}_{-0.3}$ & $1.0\,^{+0.3}_{-0.2}$ & $7.0\,^{+1.3}_{-1.3}$ & $17.4\,^{+2.8}_{-2.7}$ & $15\,^{+2.1}_{-1.8}$ & $368\,^{+51.8}_{-15.8}$ & $992\,^{+20.2}_{-28.5}$ \\
&10 & $1.0\,^{+0.5}_{-0.3}$ & $1.0\,^{+0.4}_{-0.3}$ & $5.0\,^{+1.4}_{-1.0}$ & $12.4\,^{+2.9}_{-2.2}$ & $10\,^{+1.2}_{-1.2}$ & $258\,^{+23.2}_{-14.3}$ & $649\,^{+6.4}_{-11.6}$ \\
&11 & $1.3\,^{+0.4}_{-0.3}$ & $1.3\,^{+0.4}_{-0.3}$ & $4.9\,^{+0.9}_{-0.8}$ & $12.2\,^{+1.4}_{-1.2}$ & $10\,^{+1.1}_{-1.2}$ & $212\,^{+14.0}_{-22.2}$ & $493\,^{+9.9}_{-18.7}$ \\
&12 & $1.0\,^{+0.4}_{-0.4}$ & $0.7\,^{+0.4}_{-0.3}$ & $3.5\,^{+1.6}_{-1.1}$ & $8.6\,^{+4.2}_{-2.8}$ & $8\,^{+2.8}_{-2.4}$ & $194\,^{+30.3}_{-25.4}$ & $434\,^{+10.8}_{-50.7}$ \\
&13 & $0.6\,^{+0.7}_{-0.4}$ & $0.3\,^{+0.4}_{-0.2}$ & $0.7\,^{+0.7}_{-0.7}$ & $1.4\,^{+1.2}_{-1.0}$ & $2\,^{+0.8}_{-1.1}$ & $78\,^{+25.1}_{-11.9}$ & $217\,^{+16.5}_{-35.3}$ \\
&14 & $0.3\,^{+0.4}_{-0.2}$ & $0.3\,^{+0.4}_{-0.3}$ & $-0.1\,^{+0.8}_{-0.8}$ & $-0.3\,^{+0.6}_{-0.6}$ & $3\,^{+1.6}_{-2.0}$ & $72\,^{+12.7}_{-16.9}$ & $165\,^{+17.3}_{-4.0}$ \\
&15 & $0.0\,^{+0.3}_{-0.2}$ & $0.0\,^{+0.3}_{-0.2}$ & $0.4\,^{+0.9}_{-0.8}$ & $0.9\,^{+2.0}_{-1.4}$ & $2\,^{+1.4}_{-1.4}$ & $109\,^{+15.2}_{-19.7}$ & $125\,^{+14.9}_{-1.3}$ \\
&16 & $4.1\,^{+2.3}_{-1.3}$ & $2.7\,^{+1.0}_{-0.7}$ & $22.2\,^{+10.4}_{-5.8}$ & $56.0\,^{+26.7}_{-15.2}$ & $60\,^{+23.9}_{-15.7}$ & $975\,^{+256.4}_{-114.8}$ & $2442\,^{+111.6}_{-338.8}$ \\
&17 & $3.3\,^{+1.1}_{-0.7}$ & $3.0\,^{+0.7}_{-0.7}$ & $11.7\,^{+2.6}_{-1.5}$ & $24.8\,^{+7.1}_{-3.2}$ & $42\,^{+13.2}_{-10.4}$ & $719\,^{+116.4}_{-82.9}$ & $1680\,^{+52.0}_{-90.2}$ \\
&18 & $2.8\,^{+0.7}_{-0.5}$ & $1.8\,^{+0.5}_{-0.5}$ & $11.9\,^{+1.7}_{-2.2}$ & $27.6\,^{+3.6}_{-6.3}$ & $33\,^{+3.4}_{-9.6}$ & $598\,^{+85.1}_{-77.5}$ & $1109\,^{+270.3}_{-151.8}$ \\
&19 & $2.1\,^{+1.1}_{-0.7}$ & $1.4\,^{+0.7}_{-0.4}$ & $6.0\,^{+1.6}_{-1.6}$ & $8.9\,^{+2.3}_{-2.5}$ & $13\,^{+2.1}_{-2.4}$ & $233\,^{+52.3}_{-52.9}$ & $624\,^{+5.2}_{-113.2}$ \\
&20 & $1.1\,^{+0.7}_{-0.3}$ & $1.4\,^{+0.5}_{-0.4}$ & $1.9\,^{+1.0}_{-0.8}$ & $2.0\,^{+1.1}_{-1.2}$ & $7\,^{+1.4}_{-1.3}$ & $127\,^{+15.5}_{-15.0}$ & $136\,^{+61.5}_{-7.1}$ \\
&21 & $0.2\,^{+0.4}_{-0.2}$ & $0.2\,^{+0.4}_{-0.2}$ & $0.2\,^{+0.8}_{-0.8}$ & $-1.5\,^{+1.1}_{-0.9}$ & $1\,^{+1.1}_{-1.0}$ & $43\,^{+17.4}_{-14.1}$ & $65\,^{+54.5}_{-111.5}$ \\
&22 & $0.8\,^{+0.6}_{-0.3}$ & $1.2\,^{+0.7}_{-0.5}$ & $1.2\,^{+0.9}_{-0.8}$ & $0.2\,^{+2.0}_{-1.1}$ & $2\,^{+1.6}_{-1.2}$ & $44\,^{+9.3}_{-8.1}$ & $-49\,^{+8.0}_{-0.0}$ \\
&23 & $1.3\,^{+0.4}_{-0.3}$ & $1.1\,^{+0.3}_{-0.3}$ & $6.2\,^{+1.2}_{-1.1}$ & $14.1\,^{+3.7}_{-2.5}$ & $12\,^{+3.5}_{-2.3}$ & $330\,^{+65.9}_{-55.7}$ & $1072\,^{+37.5}_{-144.9}$ \\
&24 & $1.1\,^{+0.3}_{-0.3}$ & $1.0\,^{+0.4}_{-0.2}$ & $4.7\,^{+0.8}_{-1.0}$ & $10.9\,^{+1.7}_{-1.7}$ & $9\,^{+1.2}_{-1.3}$ & $232\,^{+14.6}_{-15.4}$ & $622\,^{+19.9}_{-22.1}$ \\
&25 & $1.3\,^{+0.4}_{-0.3}$ & $1.1\,^{+0.3}_{-0.3}$ & $6.0\,^{+0.8}_{-1.4}$ & $14.6\,^{+2.4}_{-2.1}$ & $12\,^{+1.8}_{-1.8}$ & $267\,^{+31.9}_{-28.8}$ & $582\,^{+11.3}_{-10.0}$ \\
&26 & $2.1\,^{+0.7}_{-0.5}$ & $2.0\,^{+0.6}_{-0.5}$ & $6.6\,^{+1.9}_{-1.0}$ & $15.6\,^{+3.7}_{-2.0}$ & $18\,^{+2.5}_{-1.8}$ & $359\,^{+33.5}_{-38.5}$ & $728\,^{+28.3}_{-1.9}$ \\
&27 & $1.2\,^{+0.4}_{-0.3}$ & $0.9\,^{+0.4}_{-0.2}$ & $5.8\,^{+0.8}_{-0.9}$ & $14.0\,^{+1.3}_{-1.8}$ & $13\,^{+1.0}_{-1.3}$ & $322\,^{+20.6}_{-26.6}$ & $745\,^{+11.2}_{-81.0}$ \\
&28 & $1.6\,^{+0.3}_{-0.4}$ & $1.0\,^{+0.3}_{-0.2}$ & $9.7\,^{+2.0}_{-1.8}$ & $25.4\,^{+5.1}_{-4.9}$ & $22\,^{+4.1}_{-4.3}$ & $419\,^{+21.2}_{-23.2}$ & $738\,^{+7.5}_{-48.9}$ \\
&29 & $2.2\,^{+0.4}_{-0.3}$ & $1.1\,^{+0.3}_{-0.2}$ & $13.7\,^{+2.4}_{-2.1}$ & $39.1\,^{+4.8}_{-5.1}$ & $28\,^{+3.1}_{-3.9}$ & $422\,^{+27.0}_{-30.1}$ & $881\,^{+49.6}_{-3.2}$ \\
\hline
N144& 1 & $30.2\,^{+35.4}_{-11.7}$ & $40.6\,^{+40.8}_{-16.4}$ & $153.4\,^{+183.5}_{-49.0}$ & $343.5\,^{+480.9}_{-140.6}$ & $3388\,^{+15462.1}_{-1686.1}$ & $25438\,^{+34447.6}_{-7628.3}$ & $6879\,^{+1613.1}_{-401.5}$ \\
& 2 & $17.1\,^{+10.8}_{-4.7}$ & $16.1\,^{+7.6}_{-4.0}$ & $52.2\,^{+24.8}_{-12.5}$ & $114.3\,^{+68.2}_{-28.8}$ & $687\,^{+379.6}_{-107.4}$ & $10217\,^{+2895.9}_{-2311.5}$ & $9491\,^{+577.9}_{-293.9}$ \\
& 3 & $30.2\,^{+14.8}_{-9.6}$ & $25.3\,^{+10.1}_{-5.2}$ & $124.9\,^{+58.7}_{-41.2}$ & $321.2\,^{+148.3}_{-113.8}$ & $1135\,^{+995.6}_{-187.2}$ & $14521\,^{+4831.1}_{-1751.8}$ & $12412\,^{+106.0}_{-2452.4}$ \\
& 4 & $19.9\,^{+8.2}_{-4.6}$ & $15.3\,^{+5.4}_{-2.3}$ & $88.7\,^{+36.6}_{-17.7}$ & $233.0\,^{+106.3}_{-49.3}$ & $514\,^{+258.1}_{-94.1}$ & $10066\,^{+1378.3}_{-1687.3}$ & $6761\,^{+704.3}_{-246.6}$ \\
& 5 & $12.1\,^{+5.4}_{-3.8}$ & $10.3\,^{+3.8}_{-2.4}$ & $53.4\,^{+24.5}_{-15.0}$ & $136.3\,^{+65.3}_{-39.6}$ & $265\,^{+34.8}_{-41.6}$ & $4191\,^{+417.5}_{-287.1}$ & $4914\,^{+107.3}_{-54.4}$ \\
& 6 & $8.0\,^{+2.4}_{-1.5}$ & $6.5\,^{+1.2}_{-0.8}$ & $40.4\,^{+10.2}_{-5.3}$ & $104.3\,^{+27.3}_{-13.5}$ & $192\,^{+15.1}_{-21.3}$ & $4035\,^{+460.1}_{-261.6}$ & $4610\,^{+26.0}_{-55.7}$ \\
& 7 & $7.6\,^{+2.5}_{-1.4}$ & $5.8\,^{+1.7}_{-0.9}$ & $39.1\,^{+7.6}_{-6.7}$ & $98.5\,^{+19.4}_{-21.3}$ & $115\,^{+24.8}_{-18.5}$ & $2824\,^{+455.2}_{-530.0}$ & $2669\,^{+274.2}_{-81.9}$ \\
& 8 & $1.8\,^{+1.8}_{-1.0}$ & $1.5\,^{+1.1}_{-0.9}$ & $13.5\,^{+2.9}_{-4.3}$ & $29.6\,^{+7.6}_{-8.1}$ & $44\,^{+9.2}_{-8.5}$ & $1244\,^{+56.7}_{-76.2}$ & $1761\,^{+68.1}_{-23.8}$ \\
& 9 & $2.1\,^{+1.4}_{-1.0}$ & $1.6\,^{+0.8}_{-0.6}$ & $8.7\,^{+2.7}_{-2.7}$ & $21.2\,^{+6.7}_{-5.9}$ & $34\,^{+10.2}_{-9.2}$ & $996\,^{+230.7}_{-169.1}$ & $1081\,^{+41.4}_{-96.0}$ \\
&10 & $6.6\,^{+3.5}_{-2.7}$ & $6.0\,^{+2.6}_{-2.5}$ & $40.1\,^{+18.0}_{-22.8}$ & $77.2\,^{+44.6}_{-47.2}$ & $507\,^{+410.1}_{-188.4}$ & $13181\,^{+4492.3}_{-7120.1}$ & $8156\,^{+2646.3}_{-3300.4}$ \\
&11 & $1.7\,^{+0.8}_{-0.7}$ & $1.5\,^{+0.5}_{-0.4}$ & $10.5\,^{+2.2}_{-2.3}$ & $16.5\,^{+6.3}_{-4.6}$ & $103\,^{+25.2}_{-20.4}$ & $1450\,^{+68.9}_{-144.1}$ & $2821\,^{+483.0}_{-68.5}$ \\
&12 & $1.9\,^{+1.1}_{-0.6}$ & $2.5\,^{+0.6}_{-0.7}$ & $8.6\,^{+1.7}_{-1.7}$ & $17.3\,^{+3.8}_{-3.8}$ & $72\,^{+7.6}_{-6.7}$ & $1023\,^{+58.9}_{-35.5}$ & $1431\,^{+323.2}_{-12.7}$ \\
&13 & $1.2\,^{+1.2}_{-0.7}$ & $1.6\,^{+0.8}_{-0.6}$ & $2.7\,^{+2.6}_{-1.1}$ & $7.0\,^{+4.9}_{-2.7}$ & $55\,^{+5.5}_{-3.3}$ & $725\,^{+81.4}_{-43.8}$ & $888\,^{+116.8}_{-1.0}$ \\
&14 & $0.9\,^{+1.1}_{-0.6}$ & $1.1\,^{+0.7}_{-0.4}$ & $2.2\,^{+1.7}_{-1.8}$ & $4.4\,^{+3.6}_{-1.9}$ & $37\,^{+6.4}_{-2.8}$ & $437\,^{+19.9}_{-53.0}$ & $690\,^{+23.5}_{-3.9}$ \\
&15 & $0.6\,^{+0.7}_{-0.4}$ & $0.9\,^{+0.5}_{-0.3}$ & $0.7\,^{+1.7}_{-1.2}$ & $3.2\,^{+3.1}_{-1.9}$ & $30\,^{+2.1}_{-3.0}$ & $287\,^{+39.5}_{-31.7}$ & $474\,^{+41.3}_{-54.1}$ \\
&16 & $-0.5\,^{+0.5}_{-0.3}$ & $-0.2\,^{+0.4}_{-0.3}$ & $-2.1\,^{+0.9}_{-0.7}$ & $-3.0\,^{+0.8}_{-0.7}$ & $12\,^{+2.3}_{-2.6}$ & $75\,^{+20.6}_{-18.4}$ & $290\,^{+17.1}_{-15.5}$ \\
&17 & $-0.4\,^{+0.6}_{-0.4}$ & $-0.3\,^{+0.5}_{-0.3}$ & $-2.2\,^{+0.7}_{-0.8}$ & $-3.4\,^{+0.7}_{-0.8}$ & $4\,^{+1.2}_{-1.4}$ & $-27\,^{+15.1}_{-21.3}$ & $111\,^{+46.5}_{-10.8}$ \\
&18 & $10.0\,^{+2.6}_{-1.6}$ & $9.6\,^{+1.6}_{-1.1}$ & $49.4\,^{+9.9}_{-11.0}$ & $119.8\,^{+29.8}_{-30.6}$ & $502\,^{+58.4}_{-53.3}$ & $5938\,^{+836.1}_{-410.4}$ & $5769\,^{+654.0}_{-208.8}$ \\
&19 & $10.8\,^{+4.9}_{-2.3}$ & $8.9\,^{+2.8}_{-1.6}$ & $49.9\,^{+12.5}_{-8.6}$ & $130.1\,^{+30.3}_{-22.6}$ & $303\,^{+59.1}_{-45.4}$ & $3504\,^{+781.3}_{-531.5}$ & $3510\,^{+193.1}_{-119.4}$ \\
&20 & $9.6\,^{+2.9}_{-2.4}$ & $7.9\,^{+2.1}_{-1.5}$ & $34.9\,^{+8.0}_{-7.8}$ & $86.1\,^{+17.0}_{-19.3}$ & $183\,^{+15.9}_{-21.0}$ & $2009\,^{+166.5}_{-165.5}$ & $2349\,^{+134.9}_{-779.9}$ \\
&21 & $1.6\,^{+0.8}_{-0.6}$ & $1.5\,^{+0.6}_{-0.5}$ & $7.6\,^{+2.1}_{-1.7}$ & $16.4\,^{+5.2}_{-5.1}$ & $51\,^{+13.9}_{-11.6}$ & $782\,^{+229.4}_{-172.9}$ & $1315\,^{+115.2}_{-246.4}$ \\
&22 & $1.1\,^{+0.8}_{-0.5}$ & $0.8\,^{+0.6}_{-0.3}$ & $5.2\,^{+1.8}_{-1.5}$ & $13.0\,^{+4.7}_{-4.5}$ & $31\,^{+4.4}_{-5.6}$ & $501\,^{+75.8}_{-81.1}$ & $611\,^{+185.6}_{-120.2}$ \\
&23 & $0.3\,^{+0.6}_{-0.4}$ & $0.0\,^{+0.5}_{-0.3}$ & $1.9\,^{+1.3}_{-1.1}$ & $3.0\,^{+2.2}_{-1.3}$ & $10\,^{+4.7}_{-3.0}$ & $221\,^{+54.9}_{-44.3}$ & $406\,^{+27.0}_{-49.6}$ \\
&24 & $10.0\,^{+3.9}_{-2.0}$ & $10.1\,^{+3.0}_{-2.3}$ & $31.7\,^{+9.8}_{-5.9}$ & $47.3\,^{+20.1}_{-17.6}$ & $503\,^{+223.3}_{-54.9}$ & $6364\,^{+6282.2}_{-3030.7}$ & $4376\,^{+294.3}_{-555.0}$ \\
&25 & $8.9\,^{+6.1}_{-3.5}$ & $7.7\,^{+4.1}_{-3.0}$ & $15.1\,^{+6.4}_{-6.0}$ & $18.5\,^{+9.1}_{-8.7}$ & $172\,^{+17.3}_{-38.3}$ & $1544\,^{+219.9}_{-186.6}$ & $2791\,^{+105.3}_{-743.9}$ \\
&26 & $3.9\,^{+1.8}_{-1.1}$ & $3.7\,^{+1.3}_{-0.7}$ & $8.0\,^{+7.0}_{-3.3}$ & $9.5\,^{+18.5}_{-6.0}$ & $98\,^{+23.2}_{-12.2}$ & $852\,^{+200.8}_{-56.6}$ & $1111\,^{+107.4}_{-60.1}$ \\
&27 & $3.3\,^{+1.5}_{-0.9}$ & $3.1\,^{+1.0}_{-0.6}$ & $6.6\,^{+2.9}_{-1.4}$ & $8.3\,^{+5.3}_{-2.0}$ & $76\,^{+11.6}_{-8.3}$ & $748\,^{+133.6}_{-98.0}$ & $1049\,^{+2.0}_{-85.1}$ \\
&28 & $1.8\,^{+2.6}_{-0.5}$ & $1.7\,^{+1.6}_{-0.4}$ & $6.0\,^{+5.5}_{-1.8}$ & $5.9\,^{+16.0}_{-2.7}$ & $46\,^{+11.5}_{-7.7}$ & $514\,^{+125.3}_{-33.8}$ & $907\,^{+20.8}_{-25.9}$ \\
&29 & $2.2\,^{+1.3}_{-0.6}$ & $2.1\,^{+1.0}_{-0.7}$ & $7.5\,^{+2.6}_{-2.5}$ & $18.1\,^{+5.4}_{-6.0}$ & $44\,^{+6.9}_{-7.5}$ & $598\,^{+28.2}_{-52.1}$ & $1108\,^{+69.2}_{-48.0}$ \\
&30 & $3.6\,^{+1.6}_{-0.9}$ & $3.4\,^{+0.7}_{-0.6}$ & $13.5\,^{+4.3}_{-2.9}$ & $34.1\,^{+13.4}_{-8.2}$ & $59\,^{+8.2}_{-5.5}$ & $884\,^{+71.4}_{-76.3}$ & $1663\,^{+21.0}_{-160.4}$ \\
&31 & $3.0\,^{+1.2}_{-1.1}$ & $2.3\,^{+0.8}_{-0.5}$ & $8.2\,^{+4.0}_{-2.0}$ & $20.9\,^{+7.4}_{-4.8}$ & $46\,^{+6.8}_{-5.9}$ & $931\,^{+63.4}_{-57.6}$ & $1455\,^{+99.5}_{-56.6}$ \\
\enddata
\tablenotetext{a}{The flux densities are measured via 40\arcsec-diameter circular apertures and are background subtracted. Negative flux values indicate that the aperture fluxes are similar to the background fluxes.}
\end{deluxetable}

\newpage
\begin{deluxetable}{ccccccccc}
\tabletypesize{\scriptsize}
\tablewidth{0pc}
\tablecaption{Dust Properties from SED Modeling\tablenotemark{a} \label{tab_dustem} }
\tablehead{
\colhead{HII}    	&
\colhead{Subregion}    	&
\colhead{$U$\tablenotemark{b}}     	&
\colhead{$Y_{\rm PAH}/Y_{\rm DUST}$}      	&
\colhead{$Y_{\rm VSG}/Y_{\rm BG}$}     	&
\colhead{$Y_{\rm VSG}/Y_{\rm DUST}$}      	&
\colhead{$T_{\rm eq}$}     	&
\colhead{$A_{\rm V}$\tablenotemark{c}}      	\\
\colhead{Region}&
\colhead{}&
\colhead{}     	&
\colhead{($\times10^{-2}$)}     	&
\colhead{($\times10^{-2}$)}      	&
\colhead{($\times10^{-2}$)}      	&
\colhead{(K)} &
\colhead{(mag)}   	 
}
\startdata
N63&    1 & \nodata & \nodata & \nodata & \nodata & \nodata & \nodata \\
&    2 & ${9.0}{\pm3.7}$ &  ${4.6}{\pm2.0}$ &  ${34.9}{\pm20.4}$ &  ${24.7}{\pm14.7}$ &  ${26.6}^{+0.7}_{-2.4}$ &  ${0.37}{\pm0.06}$ &  \\ 
&    3 & ${10.4}{\pm1.1}$ &  ${5.3}{\pm1.2}$ &  ${22.1}{\pm5.1}$ &  ${17.1}{\pm3.9}$ &  ${27.2}^{+0.8}_{-2.5}$ &  ${0.23}{\pm0.01}$ &  \\ 
&    4 & ${15.5}{\pm1.7}$ &  ${4.8}{\pm1.9}$ &  ${25.3}{\pm6.6}$ &  ${19.2}{\pm5.0}$ &  ${29.2}^{+0.8}_{-2.7}$ &  ${0.13}{\pm0.01}$ &  \\ 
&    5 & ${8.9}{\pm0.8}$ &  ${4.5}{\pm0.8}$ &  ${13.2}{\pm3.5}$ &  ${11.1}{\pm2.9}$ &  ${26.5}^{+0.7}_{-2.4}$ &  ${0.12}{\pm0.01}$ &  \\ 
&    6 & ${5.7}{\pm0.5}$ &  ${4.8}{\pm1.1}$ &  ${8.9}{\pm2.8}$ &  ${7.8}{\pm2.4}$ &  ${24.6}^{+0.7}_{-2.2}$ &  ${0.15}{\pm0.01}$ &  \\ 
&    7 & ${5.0}{\pm1.0}$ &  ${1.2}{\pm1.0}$ &  ${5.3}{\pm3.2}$ &  ${5.0}{\pm3.0}$ &  ${24.0}^{+0.7}_{-2.2}$ &  ${0.21}{\pm0.03}$ &  \\ 
&    8 & \nodata & \nodata & \nodata & \nodata & \nodata & \\
&    9 & ${9.7}{\pm4.9}$ &  ${5.2}{\pm3.6}$ &  ${9.1}{\pm12.0}$ &  ${7.9}{\pm10.5}$ &  ${26.9}^{+0.7}_{-2.4}$ &  ${0.48}{\pm0.08}$ &  \\ 
&   10 & ${1.1}{\pm0.4}$ &  ${9.6}{\pm9.0}$ &  ${3.7}{\pm16.5}$ &  ${3.3}{\pm14.4}$ &  ${18.6}^{+0.5}_{-1.6}$ &  ${1.13}{\pm0.39}$ &  \\ 
&   11 & \nodata & \nodata & \nodata & \nodata & \nodata & \\
&   12 & \nodata & \nodata & \nodata & \nodata & \nodata & \nodata \\
&   13 & ${12.6}{\pm2.7}$ &  ${5.6}{\pm1.7}$ &  ${63.0}{\pm27.2}$ &  ${36.5}{\pm15.3}$ &  ${28.2}^{+0.8}_{-2.6}$ &  ${0.36}{\pm0.07}$ &  \\ 
&   14 & ${7.4}{\pm1.0}$ &  ${5.6}{\pm1.5}$ &  ${19.3}{\pm7.1}$ &  ${15.2}{\pm5.6}$ &  ${25.7}^{+0.7}_{-2.3}$ &  ${0.47}{\pm0.04}$ &  \\ 
&   15 & ${10.8}{\pm1.9}$ &  ${10.9}{\pm3.0}$ &  ${14.3}{\pm6.3}$ &  ${11.2}{\pm4.8}$ &  ${27.5}^{+0.8}_{-2.5}$ &  ${0.26}{\pm0.03}$ &  \\ 
&   16 & ${9.5}{\pm0.9}$ &  ${6.7}{\pm1.2}$ &  ${8.4}{\pm3.2}$ &  ${7.2}{\pm2.7}$ &  ${26.8}^{+0.7}_{-2.4}$ &  ${0.29}{\pm0.01}$ &  \\ 
&   17 & ${22.9}{\pm7.9}$ &  ${6.6}{\pm3.2}$ &  ${13.5}{\pm9.7}$ &  ${11.1}{\pm8.0}$ &  ${31.3}^{+0.9}_{-2.9}$ &  ${0.08}{\pm0.01}$ &  \\ 
&   18 & \nodata & \nodata & \nodata & \nodata & \nodata & \\
&   19 & ${4.5}{\pm0.4}$ &  ${3.7}{\pm1.1}$ &  ${22.5}{\pm5.9}$ &  ${17.7}{\pm4.7}$ &  ${23.6}^{+0.6}_{-2.1}$ &  ${0.35}{\pm0.02}$ &  \\ 
&   20 & ${5.8}{\pm0.6}$ &  ${6.4}{\pm1.1}$ &  ${13.1}{\pm2.9}$ &  ${10.8}{\pm2.4}$ &  ${24.6}^{+0.7}_{-2.2}$ &  ${0.30}{\pm0.01}$ &  \\ 
&   21 & ${3.9}{\pm0.2}$ &  ${5.6}{\pm1.0}$ &  ${6.9}{\pm1.9}$ &  ${6.1}{\pm1.7}$ &  ${23.0}^{+0.6}_{-2.1}$ &  ${0.42}{\pm0.02}$ &  \\ 
&   22 & ${5.0}{\pm0.5}$ &  ${6.6}{\pm1.6}$ &  ${11.2}{\pm3.5}$ &  ${9.4}{\pm2.9}$ &  ${24.0}^{+0.7}_{-2.2}$ &  ${0.35}{\pm0.01}$ &  \\ 
\hline
N180&    1 & ${14.6}{\pm6.6}$ &  ${5.3}{\pm3.0}$ &  ${41.4}{\pm25.4}$ &  ${27.7}{\pm17.4}$ &  ${28.9}^{+0.8}_{-2.6}$ &  ${0.83}{\pm0.16}$ &  \\ 
&    2 & ${9.2}{\pm2.2}$ &  ${4.9}{\pm1.5}$ &  ${39.8}{\pm14.5}$ &  ${27.1}{\pm9.7}$ &  ${26.7}^{+0.7}_{-2.4}$ &  ${0.91}{\pm0.12}$ &  \\ 
&    3 & ${7.7}{\pm1.1}$ &  ${6.5}{\pm1.3}$ &  ${45.3}{\pm9.5}$ &  ${29.1}{\pm6.2}$ &  ${25.9}^{+0.7}_{-2.3}$ &  ${1.02}{\pm0.08}$ &  \\ 
&    4 & ${10.8}{\pm2.3}$ &  ${5.8}{\pm2.0}$ &  ${27.0}{\pm8.9}$ &  ${20.0}{\pm6.7}$ &  ${27.4}^{+0.8}_{-2.5}$ &  ${0.85}{\pm0.08}$ &  \\ 
&    5 & ${33.5}{\pm9.2}$ &  ${1.8}{\pm1.9}$ &  ${16.7}{\pm6.5}$ &  ${14.0}{\pm5.5}$ &  ${33.5}^{+0.9}_{-3.1}$ &  ${0.33}{\pm0.03}$ &  \\ 
&    6 & ${3.4}{\pm0.5}$ &  ${10.0}{\pm2.4}$ &  ${0.2}{\pm3.2}$ &  ${0.2}{\pm2.8}$ &  ${22.5}^{+0.6}_{-2.0}$ &  ${1.66}{\pm0.14}$ &  \\ 
&    7 & ${3.3}{\pm0.1}$ &  ${9.0}{\pm0.5}$ &  ${0.0}{\pm0.0}$ &  ${0.0}{\pm0.0}$ &  ${22.4}^{+0.6}_{-2.0}$ &  ${0.96}{\pm0.03}$ &  \\ 
&    8 & ${3.7}{\pm0.3}$ &  ${12.3}{\pm2.1}$ &  ${0.0}{\pm0.0}$ &  ${0.0}{\pm0.0}$ &  ${22.8}^{+0.6}_{-2.0}$ &  ${0.74}{\pm0.04}$ &  \\ 
&    9 & ${2.5}{\pm0.1}$ &  ${9.8}{\pm1.0}$ &  ${0.0}{\pm0.0}$ &  ${0.0}{\pm0.0}$ &  ${21.2}^{+0.6}_{-1.9}$ &  ${1.02}{\pm0.06}$ &  \\ 
&   10 & ${14.5}{\pm0.5}$ &  ${0.0}{\pm0.0}$ &  ${16.2}{\pm1.2}$ &  ${14.0}{\pm1.1}$ &  ${28.9}^{+0.8}_{-2.6}$ &  ${0.21}{\pm0.01}$ &  \\ 
&   11 & ${5.1}{\pm0.3}$ &  ${0.9}{\pm2.5}$ &  ${10.4}{\pm3.2}$ &  ${9.4}{\pm2.9}$ &  ${24.1}^{+0.7}_{-2.2}$ &  ${0.36}{\pm0.02}$ &  \\ 
&   12 & ${1.2}{\pm0.0}$ &  ${10.4}{\pm1.1}$ &  ${0.0}{\pm0.0}$ &  ${0.0}{\pm0.0}$ &  ${18.8}^{+0.5}_{-1.7}$ &  ${1.62}{\pm0.12}$ &  \\ 
&   13 & ${4.1}{\pm0.3}$ &  ${6.1}{\pm0.9}$ &  ${7.1}{\pm1.9}$ &  ${6.2}{\pm1.6}$ &  ${23.2}^{+0.6}_{-2.1}$ &  ${0.42}{\pm0.01}$ &  \\ 
&   14 & ${18.0}{\pm3.2}$ &  ${6.0}{\pm1.4}$ &  ${22.5}{\pm9.0}$ &  ${17.3}{\pm6.9}$ &  ${30.0}^{+0.8}_{-2.7}$ &  ${0.60}{\pm0.07}$ &  \\ 
&   15 & ${15.6}{\pm5.6}$ &  ${5.8}{\pm2.7}$ &  ${14.8}{\pm9.5}$ &  ${12.2}{\pm7.6}$ &  ${29.3}^{+0.8}_{-2.7}$ &  ${0.47}{\pm0.11}$ &  \\ 
&   16 & ${7.4}{\pm2.6}$ &  ${6.1}{\pm2.4}$ &  ${11.8}{\pm7.6}$ &  ${9.9}{\pm6.4}$ &  ${25.7}^{+0.7}_{-2.3}$ &  ${0.38}{\pm0.06}$ &  \\ 
&   17 & ${9.0}{\pm0.9}$ &  ${5.2}{\pm1.0}$ &  ${9.5}{\pm2.6}$ &  ${8.2}{\pm2.2}$ &  ${26.6}^{+0.7}_{-2.4}$ &  ${0.21}{\pm0.02}$ &  \\ 
&   18 & ${8.0}{\pm2.9}$ &  ${5.4}{\pm3.5}$ &  ${1.8}{\pm3.8}$ &  ${1.6}{\pm3.6}$ &  ${26.0}^{+0.7}_{-2.4}$ &  ${0.11}{\pm0.04}$ &  \\ 
&   19 & ${11.0}{\pm3.7}$ &  ${0.7}{\pm2.3}$ &  ${1.7}{\pm3.9}$ &  ${1.6}{\pm3.8}$ &  ${27.5}^{+0.8}_{-2.5}$ &  ${0.03}{\pm0.01}$ &  \\ 
&   20 & ${62.6}{\pm24.7}$ &  ${10.1}{\pm6.9}$ &  ${0.0}{\pm0.0}$ &  ${0.0}{\pm0.0}$ &  ${37.4}^{+1.1}_{-3.5}$ &  ${0.01}{\pm0.01}$ &  \\ 
\hline
N70&    1 & ${5.5}{\pm0.8}$ &  ${7.8}{\pm1.5}$ &  ${1.7}{\pm2.8}$ &  ${1.6}{\pm2.5}$ &  ${24.4}^{+0.7}_{-2.2}$ &  ${0.48}{\pm0.04}$ &  \\ 
&    2 & ${9.1}{\pm1.9}$ &  ${8.4}{\pm3.6}$ &  ${6.6}{\pm5.0}$ &  ${5.7}{\pm4.3}$ &  ${26.6}^{+0.7}_{-2.4}$ &  ${0.21}{\pm0.03}$ &  \\ 
&    3 & ${5.4}{\pm1.3}$ &  ${5.9}{\pm1.7}$ &  ${15.4}{\pm12.5}$ &  ${12.6}{\pm10.2}$ &  ${24.3}^{+0.7}_{-2.2}$ &  ${0.27}{\pm0.04}$ &  \\ 
&    4 & ${13.3}{\pm3.6}$ &  ${13.5}{\pm5.0}$ &  ${5.1}{\pm5.7}$ &  ${4.2}{\pm4.7}$ &  ${28.5}^{+0.8}_{-2.6}$ &  ${0.07}{\pm0.01}$ &  \\ 
&    5 & ${6.9}{\pm0.8}$ &  ${5.0}{\pm1.9}$ &  ${8.4}{\pm3.7}$ &  ${7.3}{\pm3.2}$ &  ${25.4}^{+0.7}_{-2.3}$ &  ${0.08}{\pm0.01}$ &  \\ 
&    6 & \nodata & \nodata & \nodata & \nodata & \nodata & \nodata \\
&    7 & \nodata & \nodata & \nodata & \nodata & \nodata & \nodata \\
&    8 & \nodata & \nodata & \nodata & \nodata & \nodata & \nodata \\
&    9 & ${4.4}{\pm0.3}$ &  ${8.1}{\pm1.1}$ &  ${1.1}{\pm1.7}$ &  ${1.0}{\pm1.5}$ &  ${23.5}^{+0.6}_{-2.1}$ &  ${0.32}{\pm0.01}$ &  \\ 
&   10 & ${7.8}{\pm0.5}$ &  ${0.0}{\pm1.0}$ &  ${17.4}{\pm2.7}$ &  ${14.8}{\pm2.3}$ &  ${25.9}^{+0.7}_{-2.3}$ &  ${0.12}{\pm0.01}$ &  \\ 
&   11 & ${5.3}{\pm0.4}$ &  ${10.4}{\pm1.2}$ &  ${1.8}{\pm1.8}$ &  ${1.6}{\pm1.6}$ &  ${24.3}^{+0.7}_{-2.2}$ &  ${0.14}{\pm0.01}$ &  \\ 
&   12 & ${5.5}{\pm0.8}$ &  ${8.4}{\pm2.7}$ &  ${3.0}{\pm4.7}$ &  ${2.7}{\pm4.2}$ &  ${24.4}^{+0.7}_{-2.2}$ &  ${0.12}{\pm0.01}$ &  \\ 
&   13 & ${3.8}{\pm0.7}$ &  ${2.7}{\pm2.2}$ &  ${4.7}{\pm4.2}$ &  ${4.3}{\pm3.9}$ &  ${22.9}^{+0.6}_{-2.0}$ &  ${0.08}{\pm0.01}$ &  \\ 
&   14 & ${4.2}{\pm1.0}$ &  ${0.0}{\pm0.0}$ &  ${12.8}{\pm9.7}$ &  ${11.3}{\pm8.7}$ &  ${23.3}^{+0.6}_{-2.1}$ &  ${0.05}{\pm0.01}$ &  \\ 
&   15 & ${15.8}{\pm3.7}$ &  ${1.5}{\pm3.6}$ &  ${6.6}{\pm6.8}$ &  ${6.1}{\pm6.3}$ &  ${29.3}^{+0.8}_{-2.7}$ &  ${0.02}{\pm0.01}$ &  \\ 
&   16 & ${4.5}{\pm0.8}$ &  ${10.5}{\pm3.1}$ &  ${4.1}{\pm6.3}$ &  ${3.5}{\pm5.5}$ &  ${23.5}^{+0.6}_{-2.1}$ &  ${0.79}{\pm0.09}$ &  \\ 
&   17 & ${6.7}{\pm1.0}$ &  ${5.8}{\pm1.2}$ &  ${9.9}{\pm5.4}$ &  ${8.5}{\pm4.7}$ &  ${25.3}^{+0.7}_{-2.3}$ &  ${0.37}{\pm0.03}$ &  \\ 
&   18 & ${7.0}{\pm1.6}$ &  ${10.3}{\pm3.3}$ &  ${5.3}{\pm4.2}$ &  ${4.5}{\pm3.5}$ &  ${25.4}^{+0.7}_{-2.3}$ &  ${0.26}{\pm0.05}$ &  \\ 
&   19 & ${3.7}{\pm0.7}$ &  ${7.6}{\pm2.1}$ &  ${6.5}{\pm3.8}$ &  ${5.6}{\pm3.3}$ &  ${22.8}^{+0.6}_{-2.0}$ &  ${0.23}{\pm0.01}$ &  \\ 
&   20 & ${16.1}{\pm5.0}$ &  ${4.7}{\pm3.1}$ &  ${20.0}{\pm11.4}$ &  ${15.9}{\pm8.8}$ &  ${29.4}^{+0.8}_{-2.7}$ &  ${0.02}{\pm0.01}$ &  \\ 
&   21 & \nodata & \nodata & \nodata & \nodata & \nodata & \nodata \\
&   22 & \nodata & \nodata & \nodata & \nodata & \nodata & \nodata \\
&   23 & ${3.5}{\pm0.4}$ &  ${6.9}{\pm1.4}$ &  ${0.5}{\pm2.1}$ &  ${0.4}{\pm1.9}$ &  ${22.5}^{+0.6}_{-2.0}$ &  ${0.41}{\pm0.03}$ &  \\ 
&   24 & ${4.5}{\pm0.3}$ &  ${8.1}{\pm1.1}$ &  ${0.5}{\pm1.6}$ &  ${0.5}{\pm1.5}$ &  ${23.6}^{+0.6}_{-2.1}$ &  ${0.20}{\pm0.01}$ &  \\ 
&   25 & ${6.0}{\pm0.6}$ &  ${10.4}{\pm1.6}$ &  ${1.3}{\pm2.3}$ &  ${1.1}{\pm2.0}$ &  ${24.8}^{+0.7}_{-2.2}$ &  ${0.15}{\pm0.01}$ &  \\ 
&   26 & ${5.8}{\pm0.6}$ &  ${8.7}{\pm1.5}$ &  ${5.7}{\pm2.6}$ &  ${4.9}{\pm2.3}$ &  ${24.6}^{+0.7}_{-2.2}$ &  ${0.19}{\pm0.01}$ &  \\ 
&   27 & ${5.3}{\pm0.4}$ &  ${8.2}{\pm1.0}$ &  ${2.2}{\pm1.3}$ &  ${2.0}{\pm1.2}$ &  ${24.3}^{+0.7}_{-2.2}$ &  ${0.21}{\pm0.01}$ &  \\ 
&   28 & ${8.2}{\pm0.6}$ &  ${12.4}{\pm1.9}$ &  ${1.8}{\pm3.6}$ &  ${1.6}{\pm3.1}$ &  ${26.1}^{+0.7}_{-2.4}$ &  ${0.16}{\pm0.01}$ &  \\ 
&   29 & ${5.5}{\pm0.5}$ &  ${12.4}{\pm1.8}$ &  ${10.4}{\pm4.0}$ &  ${8.2}{\pm3.2}$ &  ${24.4}^{+0.7}_{-2.2}$ &  ${0.24}{\pm0.01}$ &  \\ 
\hline
N144& 1 & \nodata & \nodata & \nodata & \nodata & \nodata & \nodata \\
&    2 & ${20.5}{\pm10.2}$ &  ${3.9}{\pm2.4}$ &  ${32.7}{\pm25.0}$ &  ${23.7}{\pm18.3}$ &  ${30.7}^{+0.9}_{-2.8}$ &  ${0.99}{\pm0.23}$ &  \\ 
&    3 & ${25.2}{\pm10.3}$ &  ${7.6}{\pm4.4}$ &  ${35.9}{\pm29.7}$ &  ${24.4}{\pm19.8}$ &  ${31.8}^{+0.9}_{-2.9}$ &  ${1.12}{\pm0.33}$ &  \\ 
&    4 & ${46.6}{\pm14.4}$ &  ${7.4}{\pm3.1}$ &  ${13.9}{\pm9.8}$ &  ${11.3}{\pm7.9}$ &  ${35.5}^{+1.0}_{-3.3}$ &  ${0.45}{\pm0.08}$ &  \\ 
&    5 & ${13.8}{\pm1.9}$ &  ${9.2}{\pm2.7}$ &  ${15.7}{\pm6.3}$ &  ${12.3}{\pm4.9}$ &  ${28.6}^{+0.8}_{-2.6}$ &  ${0.71}{\pm0.05}$ &  \\ 
&    6 & ${18.4}{\pm2.3}$ &  ${6.5}{\pm1.2}$ &  ${11.0}{\pm2.9}$ &  ${9.3}{\pm2.5}$ &  ${30.1}^{+0.8}_{-2.7}$ &  ${0.55}{\pm0.02}$ &  \\ 
&    7 & ${23.1}{\pm6.5}$ &  ${9.8}{\pm3.2}$ &  ${4.6}{\pm4.3}$ &  ${4.0}{\pm3.7}$ &  ${31.3}^{+0.9}_{-2.9}$ &  ${0.29}{\pm0.03}$ &  \\ 
&    8 & ${11.5}{\pm1.0}$ &  ${4.6}{\pm1.3}$ &  ${11.8}{\pm3.6}$ &  ${10.0}{\pm3.1}$ &  ${27.8}^{+0.8}_{-2.5}$ &  ${0.27}{\pm0.01}$ &  \\ 
&    9 & ${17.6}{\pm5.3}$ &  ${6.0}{\pm2.3}$ &  ${6.2}{\pm5.3}$ &  ${5.5}{\pm4.7}$ &  ${29.9}^{+0.8}_{-2.7}$ &  ${0.14}{\pm0.01}$ &  \\ 
&   10  & \nodata & \nodata & \nodata & \nodata & \nodata & \nodata \\
&   11 & ${15.0}{\pm3.3}$ &  ${3.3}{\pm1.2}$ &  ${29.0}{\pm12.6}$ &  ${21.8}{\pm9.3}$ &  ${29.1}^{+0.8}_{-2.6}$ &  ${0.22}{\pm0.04}$ &  \\ 
&   12 & ${14.1}{\pm1.7}$ &  ${4.9}{\pm1.0}$ &  ${24.9}{\pm5.6}$ &  ${19.0}{\pm4.2}$ &  ${28.8}^{+0.8}_{-2.6}$ &  ${0.17}{\pm0.01}$ &  \\ 
&   13 & ${13.1}{\pm1.9}$ &  ${2.8}{\pm1.2}$ &  ${32.1}{\pm6.7}$ &  ${23.6}{\pm5.0}$ &  ${28.4}^{+0.8}_{-2.6}$ &  ${0.12}{\pm0.01}$ &  \\ 
&   14 & ${23.6}{\pm2.5}$ &  ${0.0}{\pm0.0}$ &  ${57.8}{\pm10.1}$ &  ${36.6}{\pm6.2}$ &  ${31.5}^{+0.9}_{-2.9}$ &  ${0.05}{\pm0.01}$ &  \\ 
&   15 & ${4.8}{\pm1.3}$ &  ${2.6}{\pm2.1}$ &  ${53.4}{\pm19.9}$ &  ${33.9}{\pm12.5}$ &  ${23.8}^{+0.7}_{-2.1}$ &  ${0.14}{\pm0.02}$ &  \\ 
&   16 & ${1.8}{\pm0.5}$ &  ${0.0}{\pm0.0}$ &  ${23.1}{\pm12.1}$ &  ${18.8}{\pm10.0}$ &  ${20.2}^{+0.5}_{-1.8}$ &  ${0.18}{\pm0.02}$ &  \\ 
&   17 & \nodata & \nodata & \nodata & \nodata & \nodata & \nodata \\
&   18 & ${17.4}{\pm4.1}$ &  ${7.1}{\pm2.2}$ &  ${42.5}{\pm15.4}$ &  ${27.7}{\pm9.9}$ &  ${29.8}^{+0.8}_{-2.7}$ &  ${0.67}{\pm0.09}$ &  \\ 
&   19 & ${17.1}{\pm6.1}$ &  ${11.6}{\pm4.7}$ &  ${31.3}{\pm17.7}$ &  ${21.1}{\pm12.0}$ &  ${29.7}^{+0.8}_{-2.7}$ &  ${0.43}{\pm0.07}$ &  \\ 
&   20 & ${12.2}{\pm3.3}$ &  ${12.8}{\pm4.2}$ &  ${29.3}{\pm12.8}$ &  ${19.8}{\pm8.2}$ &  ${28.0}^{+0.8}_{-2.5}$ &  ${0.36}{\pm0.06}$ &  \\ 
&   21 & ${6.4}{\pm2.3}$ &  ${5.6}{\pm2.5}$ &  ${18.2}{\pm12.0}$ &  ${14.5}{\pm9.6}$ &  ${25.1}^{+0.7}_{-2.3}$ &  ${0.32}{\pm0.06}$ &  \\ 
&   22 & ${12.3}{\pm5.0}$ &  ${7.3}{\pm4.1}$ &  ${18.3}{\pm13.1}$ &  ${14.4}{\pm9.8}$ &  ${28.1}^{+0.8}_{-2.5}$ &  ${0.09}{\pm0.03}$ &  \\ 
&   23 & ${6.1}{\pm1.7}$ &  ${3.1}{\pm1.9}$ &  ${13.0}{\pm8.8}$ &  ${11.1}{\pm7.6}$ &  ${24.8}^{+0.7}_{-2.2}$ &  ${0.10}{\pm0.01}$ &  \\ 
&   24  & \nodata & \nodata & \nodata & \nodata & \nodata & \nodata \\
&   25 & ${7.9}{\pm2.3}$ &  ${3.4}{\pm1.7}$ &  ${39.9}{\pm18.4}$ &  ${27.5}{\pm12.5}$ &  ${26.0}^{+0.7}_{-2.3}$ &  ${0.49}{\pm0.09}$ &  \\ 
&   26 & ${18.9}{\pm5.4}$ &  ${4.4}{\pm3.2}$ &  ${47.4}{\pm23.2}$ &  ${30.7}{\pm14.6}$ &  ${30.3}^{+0.8}_{-2.8}$ &  ${0.10}{\pm0.02}$ &  \\ 
&   27 & ${20.8}{\pm4.2}$ &  ${2.7}{\pm1.0}$ &  ${29.7}{\pm9.4}$ &  ${22.2}{\pm7.1}$ &  ${30.8}^{+0.9}_{-2.8}$ &  ${0.11}{\pm0.01}$ &  \\ 
&   28 & ${19.6}{\pm3.2}$ &  ${1.3}{\pm2.1}$ &  ${29.4}{\pm9.7}$ &  ${22.5}{\pm7.4}$ &  ${30.5}^{+0.8}_{-2.8}$ &  ${0.09}{\pm0.01}$ &  \\ 
&   29 & ${5.3}{\pm0.6}$ &  ${7.2}{\pm1.9}$ &  ${18.1}{\pm6.9}$ &  ${14.2}{\pm5.4}$ &  ${24.2}^{+0.7}_{-2.2}$ &  ${0.31}{\pm0.03}$ &  \\ 
&   30 & ${10.6}{\pm1.3}$ &  ${6.4}{\pm1.9}$ &  ${13.6}{\pm4.2}$ &  ${11.2}{\pm3.5}$ &  ${27.4}^{+0.8}_{-2.5}$ &  ${0.25}{\pm0.02}$ &  \\ 
&   31 & ${8.4}{\pm0.9}$ &  ${5.5}{\pm1.4}$ &  ${11.1}{\pm3.7}$ &  ${9.5}{\pm3.1}$ &  ${26.3}^{+0.7}_{-2.4}$ &  ${0.29}{\pm0.02}$ &  \\ 
\enddata
\tablenotetext{a}{Using the Levenberg--Marquardt minimization method implemented in
IDL (\url{http://www.physics.wisc.edu/~craigm/idl/fitting.html}).}
\tablenotetext{b}{ISRF scaled from the solar neighborhood ISRF described in \citet{Mathis83}.}
\tablenotetext{c}{The errors in the fitted $A_V$ is approximated to be at least 0.01.}
\end{deluxetable}

\end{document}